\documentclass[11pt]{article}

\usepackage {ucs}
\usepackage [utf8x]{inputenc}
\usepackage {graphicx,wrapfig}
\usepackage {amsfonts,amsmath,amsthm}%,amssymb}
\usepackage {epstopdf} 						% to automatically transform .eps figs to .pdf
\usepackage {multicol}						% to locally have two columns
\usepackage {datetime} 						% date and time in latex
\usepackage {subfigure}						% to have enumerated subfigures in one figure
\usepackage {parskip}						% no indentation and space between paragraphs
\usepackage {pgfgantt}						% to create the pgf graphs
\usepackage {setspace}						% to set the vertical spacing 
\usepackage {enumitem}						% to set spaces in lists
\usepackage {eurosym} 						%to include the euro sign
\usepackage {textgreek}						% to include greek letters in normal text
\usepackage[normalem]{ulem}

\newcommand {\pd} {\partial}
\newcommand {\R} {\mathbb R}

\renewcommand {\a}{\alpha}
\renewcommand {\b}{\beta}

\renewcommand {\(}{\left (}
\renewcommand {\)}{\right )}
\renewcommand{\vec}{\mathbf}

\def\redd#1{{\color{red}{#1}}}
\def\subtext#1{\text{{#1}}}

\theoremstyle{remark}       %this style exists in the amsthm package
	\newtheorem{remark}{\textbf{Remark}}[section]
	\newtheorem{experiment}{\textbf{Experiment}}[section]

\title{
	A Hybrid Multiscale Model for Cancer Invasion of the Extracellular Matrix
	}

\author{
	Nikolaos Sfakianakis\thanks{
		Institute of Applied Mathematics, Faculty of Mathematics and Informatics, Heidelberg University, Heidelberg 69120, Germany,\hfill{\tt sfakiana@math.uni-heidelberg.de}	
	}
	\and
	Anotida Madzvamuse\thanks{
		Department of Mathematics, School of Mathematical and Physical Sciences, University of Sussex, Brighton BN1 9QH, UK, \hfill{\tt A.Madzvamuse@sussex.ac.uk} 
	}
	\and
	Mark A.J. Chaplain \thanks{
		Mathematical Institute, School of Mathematics and Statistics, University of St Andrews, St Andrews KY16 9SS, Scotland, \hfill{\tt majc@st-andrews.ac.uk} 
}
}

\newcommand{\blue}[1]{\textcolor{blue}{#1}}

\begin{document}
\maketitle
\thispagestyle{empty}

\begin{abstract}
The ability to locally degrade the \textit{extracellular matrix} (ECM) and interact with the tumour microenvironment is a key process distinguishing cancer from normal cells,  and is a critical step in the metastatic spread of the tumour. The invasion of the surrounding tissue involves the coordinated action between cancer cells, the ECM, the \textit{matrix degrading enzymes}, and the \textit{epithelial-to-mesenchymal transition} (EMT). This is a regulatory process through which \textit{epithelial cells} (ECs) acquire mesenchymal characteristics and transform to \textit{mesenchymal-like cells} (MCs). In this paper, we present a new mathematical model which describes the transition from a collective invasion strategy for the ECs to an individual invasion strategy for the MCs. We achieve this by formulating a coupled hybrid system consisting of \textit{partial} and \textit{stochastic differential equations} that describe the evolution of the ECs and the MCs, respectively. This approach allows one to reproduce in a very natural way fundamental qualitative features of the current biomedical understanding of cancer invasion that are not easily captured by classical modelling approaches, for example, the invasion of the ECM by self-generated gradients and the appearance of EC invasion islands outside of the main body of the tumour.
\end{abstract}


%%%%%%%%%%%%%%%%%%%%
%%%%%%%%%%%%%%%%%%%%

%----------------------------------------------------------------------------------------
\section{Introduction}

Identified as one of the \textit{hallmarks of cancer},  \cite{Hanahan.2000,Hanahan.2011}, cancer invasion (and {the} subsequent metastasis) is a complex process involving interactions between cancer cells and the extracellular matrix (the {tumour} \textit{microenvironment}) facilitated by matrix degrading enzymes. By its nature, the invasion involves the development of and changes to cell-cell and cell-matrix adhesion processes.  Broadly speaking, during the progression to full malignancy, cancer cells reduce their cell-cell adhesions and gain cell-matrix adhesions (controlled by adhesion molecules such as \textit{cadherin}). Coupled with cell migration and proliferation, this enables the local spread of cancer cells into the surrounding tissue. Any encounter with blood or lymphatic vessels in the microenvironment initiates the spread of the cancer to secondary locations in the host i.e. metastasis, which accounts for over 90\% of deaths due to cancer, \cite{Mehlen.2006,Weigelt.2005}. 

\begin{figure}
	\centering
	\includegraphics[width=0.7\linewidth]{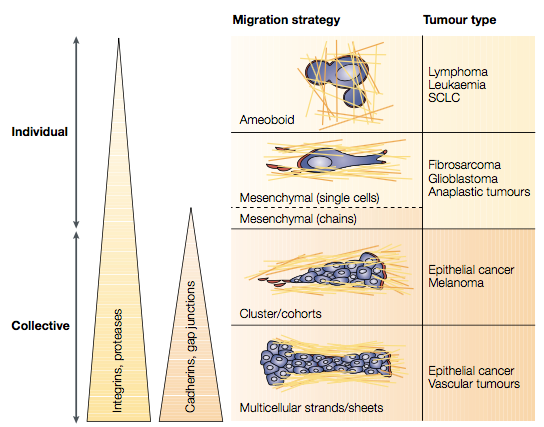} 
	\caption{``Diversity of cancer invasion''.  {Classification of the various migration and invasion strategies and corresponding types of tumour. As the complexity of the tumour increases, so do the expressions of cell-matrix and cell-adhesion molecules (\textit{integrins} and \textit{cadherins}) and the characterization of the invasion as individual or collective.} Figure adopted from \cite{Friedl.2003}. \redd{(PERMISSION REQUESTED)}}
	\label{fig:friedl1}
\end{figure}

\begin{figure}
	\centering
	\includegraphics[width=0.7\linewidth]{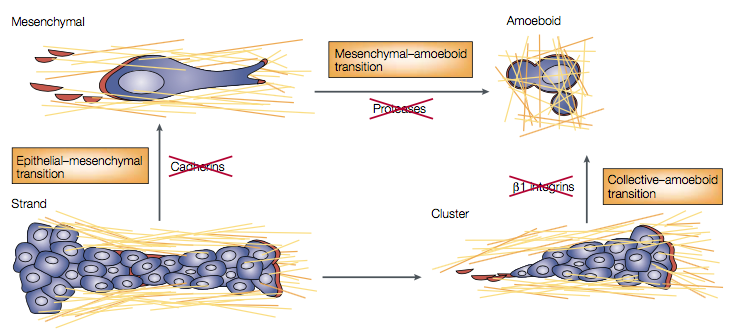} 
	\caption{``Plasticity of cancer invasion''. {The character of cancer cell migration changes from collective to individual, following the loss of the cadherin or $\beta$1 integrin function. The corresponding cellular transition programmes are conditionally reversible, leading to --in later location within the organism-- metastases.} Figure from \cite{Friedl.2003}. \redd{(PERMISSION REQUESTED)}}
	\label{fig:friedl2}
\end{figure}

Having been studied in some detail for the past 10-15 years, it has become clear that cancer invasion has a certain degree of diversity in its migratory mechanisms and a degree of plasticity in cellular behaviour and properties. The diversity of cancer invasion mechanisms is illustrated schematically in Fig. \ref{fig:friedl1} and the plasticity in  \ref{fig:friedl2}. Both  invasions are extensively discussed in \cite{Friedl.2003}. Still, cancer invasion can be broadly classified into two main groups, each differing in the behaviour of how the cells migrate ---individual versus collective migration--- and how these are controlled by different intra-cellular molecular programmes. Accordingly, cancer invasion can be characterised as \textit{epithelial} or \textit{collective invasion} whereby clusters or sheets of connected cells move en masse, or as \textit{mesenchymal} or \textit{individual invasion} whereby single cancer cells or small numbers of cancer cells actively invade the microenvironment. 

The plasticity of cancer invasion mechanisms is illustrated schematically in Fig. \ref{fig:friedl2}{, see also \cite{Friedl.2003}.} Cancer cells may transition back and forth between the two different invasion mechanisms during the invasion process as they penetrate into the surrounding tissue. The transition process between the collective and individual invasion is largely controlled by varying the expression levels of molecules such as integrins, proteases, and cadherins and varying cell-cell communication via gap junctions. This process is known as the \textit{epithelial-mesenchymal transition} (EMT), whereas the opposite process is known as \textit{mesenchymal-epithelial transition} (MET). 

An alternative invasion mechanism also exists ---\textit{amoeboid invasion}--- whereby individual cells exhibit morphological plasticity and develop the ability to squeeze through gaps in the \textit{extracellular matrix} (ECM), rather than modify/degrade the ECM via matrix degrading enzymes, e.g. {\textit{urokinase-type plasminogen activator}} (uPA), {\textit{matrix metalloproteases}} (MMPs), \cite{Madsen.2010,Sabeh.2009}. 

Cancer invasion has also been the focus of mathematical modelling over the past twenty years or so, beginning with the work of \cite{Gatenby.1996}. Since then, many different models and approaches have been formulated, some taking an individual-based (or agent-based) approach e.g. 
\cite{Ramis.2008, Ramis.2008b, Hatzikirou.2010, Wang.2013, Ramis.2015}, 
others adopting a continuum approach using systems of partial differential equations e.g.
\cite{Chaplain.1996,  Preziosi.2003, Chaplain.2005, Chaplain.2006, Andasari.2011, Domschke.2014, Deakin.2013, Painter.2013, Sfakianakis.2016, Sfakianakis.2016b, Surulescu.2017, Trucu.2017},
while others have adopted a hybrid continuum-discrete approach, e.g.
\cite{Chaplain.2000, Anderson.2005}.

An individual-based approach has the advantage of being able to focus on single cells (and usually the adhesive forces generated) and is more accurate at smaller scales, while a continuum approach has the advantage of being able to capture larger scale phenomena, perhaps better modelling collective invasion and is more accurate at large scales (e.g. tissue scale). 

Here, we present a different mathematical modelling approach to cancer invasion that explicitly models the transition from collective to individual invasion, and vice versa, by formulating a new \textit{multiscale framework} approach to the problem. In particular, we describe (cancer) \textit{epithelial cells} (ECs) by a \textit{density distribution} and their spatiotemporal evolution by a \textit{macroscopic deterministic} model, whereas (cancer) \textit{mesenchymal cells} (MCs) are modelled by an \textit{atomistic} approach, and their spatiotemporal evolution by an \textit{individual stochastic model}. In this \textit{hybrid} approach we also include the coupling dynamics between the two cell types and their corresponding descriptions.

The macroscopic sub-model for the evolution of the ECs includes primary processes such as \textit{diffusion} and \textit{proliferation} of the ECs and the \textit{de-differentiation} and \textit{differentiation} of the MCs to ECs. It also includes some basic processes of the MMPs and ECM. The MMPs are assumed to be produced by the cancer cells, to diffuse in the extracellular environment, and to degrade the ECM. The ECM, in its turn, is assumed to be non-uniform and is not remodelled. On the other hand, the MCs are described as a finite set of \textit{particles} that represent isolated cells or small cell aggregates. This representation is motivated by the small number in which the MCs appear in the tumour. For the spatiotemporal evolution of the particles we assume that they obey an \textit{individual stochastic model} that includes \textit{haptotaxis} and \textit{random motion} for each particle. The total number of particles in the system varies according to the EMT and MET processes between the ECs and MCs. 

The triggering mechanisms of the EMT and MET are not specifically addressed in this work. Instead, we follow a simplified approach and assume that the EMT occurs randomly over the ECs with a given constant probability over space and time. In a similar way, we assume that every MC particle undergoes MET randomly with a given probability rate.  Similarly, the description and modelling of specific mechanisms of the ECM dynamics also fall outside the scope of this paper. Nevertheless, we assume that the ECM is represented by the density of the \textit{collagen macromolecules}, and is therefore modelled as a spatially non-uniform, immovable component of the system. Finally, we assume that the ECM is degraded by the combined function of the cancer cells MMPs  complex and, for simplicity, that no matrix reconstruction takes place.

%==================================

%There are many biological processes that couple the ECs and the MCs; {in order to be consistent} with the dynamics that are typically considered in the macroscopic cancer invasion models, we restrict the coupling of the two cancer cell types in the EMT and MET \comm{A: a figure of some sort. N: are the figure that Mark included ok? do we need something more} { to the proliferation of the ECs}. We assume no further interactions between the cancer cells of each family. For the coupling between the ECs and MCs, we identify the two cell types as two distinct \textit{phases} and devise a basic \textit{phase transition process} responsible for the \textit{densities-to-particles} transformation and vice-versa. 

%-------------------------------------------------------------------------------------------
\section{Model derivation and interactions between the phases}\label{sec:models}

In this section we present the main components of the model, the two types (or \textit{phases}) of the cancer cells, their properties and interactions. The model we propose is a hybrid amalgam of the two phases of the cancer cells that are described by a continuum density for the DCs and a collection of discrete particles for the ECs. For the sake of clarity of presentation, the development of the model and the corresponding techniques are constrained to the two (spatial) dimensional case only.

Therefore, our paper is structured as follows. In Section \ref{sec:density} we describe the \textit{continuum density submodel} of the problem. This is a macroscopic deterministic model that addresses the spatiotemporal evolution of the densities of the ECs, ECM, and the MMPs. The participation of the DCs in this submodel is merely implicit, i.e. the macroscopic model does not dictate their time evolution. We then introduce the \textit{discrete particle} submodel of the problem in Section \ref{sec:particles}. This submodel refers to the MCs as particles and is responsible for the time evolution. In Section \ref{sec:part-dens-trans} we describe the transitions between the two phases of the cancer cells. We address the way the MC particles are substantiated from their density formulation via a \textit{density-to-particles} process, and how they transition back to density via an opposite \textit{particles-to-density} procedure. In Section \ref{sec:hybrid} we present the combined spatiotemporal evolution of the two phases under the prism of the EMT and MET processes. We moreover address separately the EMT and MET processes as well as their influence on the phases of the two types cancer cells. 

\subsection{Density formulation}\label{sec:density}
From a \textit{macroscopic deterministic} scale, we follow the seminal works of  \cite{Liotta.1977, Gatenby.1996, Chaplain.2000,  Chaplain.1999} and describe the ECs, MMPs, and ECM by their densities. As the MCs are primarily described as particles, the model for their evolution is derived in Section 2.2. Still, the MCs appear also in the density formulation as they directly affect the ECs, MMPs, and the ECM. Since the focus of this paper is on the combination of the two types/phases of the cancer cells rather than on the biological applications of the model, we consider only the very basic biological dynamics. More detailed and cancer-type specific models will be considered in follow up works. Indeed, we mainly assume that the ECs are transformed to MCs and vice-versa via the MET and EMT processes, and that they proliferate by following a \textit{logistic volume-filling} constraint as they compete for free space and resources with each other and with the MCs and the ECM. Furthermore, we assume that the ECs diffuse in the environment; although this process is expected to be very slow.

To proceed we denote by $\Omega\subset \R^2$ the Lipschitz domain of study, and by $c^\a(\vec x, t)$, $c^\b(\vec x, t)$, $m(\vec x, t)$, and  $v(\vec x, t)$, $\vec x\in \Omega$ and $t\geq 0$ the densities of the ECs, MCs, MMPs, and the ECM respectively. From here onwards, we denote by the superscripts $\alpha$ and $\beta$ the two types (or phases) of cancer cells, the ECs and MCs, respectively.

It follows therefore that the equation that controls the evolution of the ECs reads 
\begin{subequations}
\begin{align}\label{eq:dcc}
	\frac{\pd}{\pd t} c^\a(\vec x, t) = \underbrace{D_\a\Delta c^\a(\vec x, t)}_{\subtext{diffusion}} &-\underbrace{\mu_\a^\text{EMT}(\vec x,t) c^\a(\vec x, t)}_{\subtext{EMT}} +\underbrace{\mu_\b^\text{MET}(\vec x,t) c^\b(\vec x, t)}_{\subtext{MET}}\nonumber \\
	&+ \underbrace{\rho^\a_c c^\a(\vec x, t)\(1-c^\a(\vec x, t)-c^\b(\vec x, t) - v(\vec x, t)\)}_{\subtext{proliferation}},
\end{align}
where $\mu_\a^\text{EMT}(\vec x,t) = \mu_\a \mathcal X_{\mathcal A(t)}(\vec x)$, $\mu_\b^\text{MET}(\vec x,t) = \mu_\b \mathcal X_{\mathcal B(t)}(\vec x)$, {with} $\mathcal A(t), \mathcal B(t)\subset \Omega$, and $D_\a,\, \mu_\a,\, \mu_\b, \rho^\a_c\geq 0$.   

As previously noted, the MCs are described by their particle formulation ---which we present in Section \ref{sec:particles}--- and the corresponding evolutionary equations. The MCs participate also in \eqref{eq:dcc} via their density $c^\b$ after having undergone a {specific} \textit{particle-to-density} transformation; this is discussed in Section \ref{sec:part-dens-trans}.  

The triggering mechanisms of EMT and MET are not the focus of this work. We instead assume a simplified approach where EMT occurs in a randomly chosen domain, denoted by $\mathcal A(t)\subset \Omega$ in \eqref{eq:dcc}, at a constant rate $\mu_\a$. In a similar way, we assume that the MET occurs randomly at every particle; this gives rise in a natural way to the domain $\mathcal B$ and the MET rate $\mu_{\mathcal B}$, see also Sections \ref{sec:particles} and \ref{sec:part-dens-trans} for further details.

Both types of cancer cells, ECs and MCs, produce MMPs, which in turn diffuse in the environment (molecular diffusion) and decay with a constant rate satisfy:
\begin{equation}\label{eq:mmp}
	\frac{\pd}{\pd t} m(\vec x, t) = \underbrace{D_m\Delta m(\vec x, t)}_\subtext{diffusion} +\underbrace{\rho_m^\a c^\a(\vec x, t)+\rho_m^\b c^\b(\vec x, t)}_\subtext{production} - \underbrace{\lambda_m m(\vec x, t)}_\subtext{decay},
\end{equation}
with $D_m,\ \rho_m^\a,\ \rho_m^\b, \ \lambda_m\geq 0$ constants. 

The ECM is assumed to be an immovable component of the system that neither diffuses nor translocates. It is assumed to be non-uniform and to be degraded by the action of the MMP-cancer cell compound. No reconstruction of the matrix is assumed. Hence, the evolution equation of the ECM is given by
\begin{equation}\label{eq:ecm}
\frac{\pd}{\pd t} v(\vec x,t) = - \underbrace{\(\lambda_v^\a c^\a(\vec x,t) + \lambda_v^\b c^\b(\vec x,t)\) m(\vec x, t)v(\vec x,t)}_\subtext{degradation},
\end{equation}
with $\lambda_v^\a,\ \lambda_v^\b \geq 0$ constants.
\end{subequations}
%The system \eqref{eq:dcc}-\eqref{eq:ecm} is equipped with properly chosen initial conditions \remm{the smoothness of which lies outside the scope of this paper.}

The \textit{(advection-)reaction-diffusion}\footnote{In the general case, \eqref{eq:dcc}--\eqref{eq:ecm} could include advection as well} (A-)RD system \eqref{eq:dcc}--\eqref{eq:ecm} can also be written in a more convenient {matrix-vector compact form} for the numerical treatment formulation, see also Appendix \ref{sec:FV}. In particular, using the notation 
$$\vec w(\vec x,t)=\big(c^\a(\vec x,t), m(\vec x,t), v(\vec x,t)\big)^T,$$
\eqref{eq:dcc}--\eqref{eq:ecm} read
\begin{equation}\label{eq:w.model}
	\vec w_t (\vec x,t)= D(\vec w(\vec x,t)) + R(\vec w(\vec x,t)),
\end{equation}
where
\[
	D(\vec w) = \begin{pmatrix} D_\a\Delta c^\a\\ \ D_m\Delta m\\ \ 0 \end{pmatrix}\ \text{and}\  
	R(\vec w) = \begin{pmatrix} 
		-\mu_\a^\text{EMT} c^\a + \mu_\b^\text{MET} c^\b + \rho^\a_c c^\a \(1 - c^\a - c^\b - v\)\\
		\rho_m^\a c^\a+\rho_m^\b c^\b - \lambda_m m\\
		- \(\lambda_v^\a c^\a + \lambda_v^\b c^\b\) mv
	\end{pmatrix}
\]
denote the \textit{diffusion} and \textit{reaction} operators, respectively. In the more general case where chemotaxis or haptotaxis are considered, the corresponding formulation should also include an \textit{advection} operator.

Clearly, cancer invasion models of the form \eqref{eq:w.model} are mere simplifications of the biological reality; they are also quite simple in mathematical structure. Nevertheless, their analytical and numerical investigations are challenging, see for example \cite{Andasari.2011, Sfakianakis.2017, Sfakianakis.2016, Czochra.2010, Winkler.2014}. One of the reasons for this, is their mixed nature, i.e. the ECs and MMPs obey \textit{partial differential equations} (PDEs) with respect to time and space, whereas the ECM  {obeys} an \textit{ordinary differential equation} (ODE) with respect to time for every point in space.

%---------------------------------------------------------------------------------------------
\subsection{Particle formulation}\label{sec:particles}
We are motivated in this description by methods and techniques that have been used previously in other scientific fields. One such example is the classical \textit{particle-in-cell} (PIC) {method which was first proposed} in \cite{Harlow.1955} and used among others in plasma physics. A second example is be the \textit{smoothed-particle hydrodynamics} (SPH) {method} used in astrophysics and ballistics (see \cite{Gingold.1977} for example). The stochastic nature of the ODEs that the particles obey is motivated by the seminal work of \cite{Stratonovich.1966}. For the combination of the two cancer cell formulations we are inspired by \cite{Blanc.2007, Kitanidis.1994, Makridakis.2014, Thompson.1992}.

In view of the above, we describe the MCs as a system of $N$ particles that are indexed by $p\in P=\{1,\dots,N\}$, and account for their positions $\vec x_p(t)\in\R^2$ and masses $m_p(t)\geq 0$. We allow for their number to vary in time and so we set $N=N(t)\in \mathbb N$.

The mass distribution of such system of particles, $\{(\vec x_p , m_p), \;p\in P\}$, is given by 
\begin{equation}\label{eq:part.dirac}
\tilde{\tilde c} (\vec x, t)=\sum_{p\in P}m_p(t) \delta(\vec x -\vec x_p(t))
\end{equation}
where $\delta(\cdot-x_p(t))$ represents the Dirac distribution centred at $\vec x_p \in \R^2$. Clearly \eqref{eq:part.dirac} is not a function so we consider a \textit{kernel} $\zeta$ and re-define the \textit{mass distribution} of the particles $\{(\vec x_p , m_p), \; p\in P\}$ as 
\begin{align}
	\tilde c (\vec x, t)
	&= \int_{\Omega} \tilde{\tilde c}(\vec x',t)\zeta (\vec x-\vec x') d\vec x'
	\stackrel{\eqref{eq:part.dirac}}{=} \sum_{p\in P}m_p(t) \zeta(\vec x -\vec x_p(t)).\label{eq:part.better}
\end{align}

The function $\zeta$ does not have to be smooth and to simplify the rest of this work we choose it to be the characteristic function of the rectangle $K_0$ that is \textit{centred} at the origin $\vec 0\in \R^2$ 
\begin{equation}\label{eq:zeta}
	\zeta(\vec x) = \mathcal X_{K_0}(\vec x), \quad \vec x\in \R^2.
\end{equation}
The choice of $K_0$ (shape, size, and location) is justified in Sections \ref{sec:part.inter} and \ref{sec:simulations}.

%------------------------------------------------------------------------------------------------------------
\subsubsection{Interactions between particles}\label{sec:part.inter}
We understand the particles as isolated cancer cells or cancer-cell aggregates of similar size and masses. To maintain similar masses, we split and merge the particles according to their mass and position. In particular, when the particles represent an isolated cancer cell, we set $m_\text{ref}$ to be the reference mass of one cell and $K_0$ its (two-dimensional) size\footnote{We consider typically the physiological parameters of the HeLa cells as reference}, and proceed as follows:

\begin{description}
	\item[Splitting.] A particle $(\vec x_p, m_p)$ with mass $m_p>\frac{4}{3}\,m_\text{ref}$ is \textit{split} into two particles $(\vec x_p^1,m_p^1)$, $(\vec x_p^2,m_p^2)$ of the same position $\vec x_p^1=\vec x_p^2=\vec x_p$ and mass $m_p^1=m_p^2=\frac{1}{2}\,m_p$. From that moment onwards, these two particles are considered different from each other.

	\item[Merging.] A \textit{small particle} $(\vec x_p, m_p)$ with mass $m_p < \frac{2}{3}\,m_\text{ref}$ is \textit{merged} with another small particle $(\vec x_q, m_q)$ if they are close to each other i.e. 
	$$\|\vec x_p -\vec x_q\|<\text{diam} (K_0),$$
	{where $\|\cdot\|$ describes the two-dimensional Euclidean norm}. The resulting particle is set to have the cumulative mass of the two particles and to be located at their (combined) centre of mass
	\begin{equation}\label{eq:merged}
		\(\frac{m_p\vec x_p + m_q\vec x_q}{m_p+m_q},m_p+m_q\).
	\end{equation}
\end{description}

Given that the distance between the particles is sufficiently small, iterations of the \textit{merging} and \textit{splitting} processes lead to particles with masses $m_p\in[\frac{2}{3}\,m_\text{ref},\frac{4}{3}\,m_\text{ref}]$.

Besides the merging and splitting procedures, we do not consider other processes that alter the masses of the particles. Moreover, we do not include any further interactions between the particles in this work as we try to be consistent with the dynamics that are usually assumed by macroscopic deterministic models, such as \eqref{eq:dcc}--\eqref{eq:ecm}.

%--------------------------------------------------------------------------------------------------------------
\subsubsection{Time evolution of particles}
We assume that the particles perform a \textit{biased random motion} that is comprised of two independent processes: a \textit{directed-motion} part that represents the \textit{haptotactic} response of the cells to gradients of the ECM-bound adhesion sites, and a \textit{random/stochastic-motion} part that describes the undirected \textit{kinesis} of the cells as they sense the surrounding environment. We reproduce this way, at the particle level, the diffusion and -taxis dynamics prescribed by the \textit{macroscopic deterministic} cancer invasion models, see e.g. \cite{Chaplain.2000}.

We understand this complex phenomenon as a \textit{geometric Brownian motion}. In the general case, the corresponding \textit{stochastic differential equation} (SDE) that it follows, would have the form 
\begin{equation}\label{eq:brownian.1}
	d\vec X^p_t = \mu(\vec X^p_t,t) dt +\sigma(\vec X^p_t,t) d\vec W^p_t,\quad \text{for} \quad p\in P,
 \end{equation}
where $\vec X^p_t$ represents the position of the particles in physical space (here $\R^2$), and $\vec W^p_t$ a Wiener process. Here, $\mu$ and $\sigma$ are the \textit{drift} and \textit{diffusion} coefficients that encode the assumptions made on the directed and random parts of the motion of the particles. Clearly, if more complex dynamics and interactions between the particles and/or the environment are assumed, the SDE \eqref{eq:brownian.1} should be adjusted accordingly.

For the needs of this paper, we discretise \eqref{eq:brownian.1} by an \textit{It\^o-type explicit Euler-Maruyama particle motion scheme}: 
\begin{equation}\label{eq:part.loc}
	\vec x^p_{t+\tau} =\vec  x^p_t +\vec A(\vec x^p_t) \tau + \vec B(\vec x^p_t)\cdot \vec Z^p \sqrt{\tau},\quad \text{for} \quad p\in P\,,
\end{equation}
cf. \cite{Kitanidis.1994, Kloeden.1992} and Appendix \ref{sec:brownian}. Here, $\tau >0$ is the timestep of the scheme, and $\vec A$ denotes the \textit{advection operator}
\begin{equation}\label{eq:adjusted.adv}
	\vec A=\vec v +\nabla\cdot \vec D,
\end{equation}
that encodes the \textit{advection velocity} $\vec v$ adjusted by the \textit{drift term} $\vec D$. This particular choice of $\vec A$ is made so that \eqref{eq:part.loc} converges, in the many-particle limit $N\rightarrow \infty$, to the desired reaction-less version of the (A-)RD system \eqref{eq:dcc}--\eqref{eq:ecm}. The more intuitive scheme with $\vec A=\vec v$ would not converge in the case of a non-constant diffusion tensor $\vec D$; we refer to \cite{Arnold.1974, Kitanidis.1994, Raviart.1986, Stratonovich.1966, Thompson.1992} for the proofs of these claims and further discussions. 

In comparison to the usual macroscopic cancer invasion models, the advection velocity $\vec v$ in \eqref{eq:adjusted.adv} corresponds to the advection/-taxis {term}. The square matrix $\vec B$ is related to the diffusion tensor $\vec D$ by
\begin{equation}\label{eq:B}
	\vec B\cdot \vec B^T=2\vec D. 
\end{equation}
The typical Laplace operator $\mathcal L u = {d}\Delta u$, would correspond here to a diagonal $\vec D$ with inputs {$d$}. Moreover, in \eqref{eq:part.loc}, $\vec Z^p$ is a vector of normally distributed values of zero mean and unit variance. 
 	
%The discrete approach \eqref{eq:part.loc} serves also as an explicit in-time discretization of the SDE \eqref{eq:brownian.1} (in the case of a single particle) for a particualr choice of drift and diffusion coefficients, see also Appendix \ref{sec:brownian}. 
 
%-----------------------------------------------------------------------------
\subsubsection*{Modelling reactions}
Although the MCs participate in several reaction processes (such as the EMT, MET, the proliferation of the ECs, the production of MMPs, and the degradation of the ECM), the particle motion scheme \eqref{eq:part.loc} does not include any reaction terms. We account for them in the following way:

Some of the MC particles undergo MET to ECs, and subsequently are transformed to density via the \textit{particle-to-density} operator that will be introduced in Section \ref{sec:part-dens-trans}. These MCs are removed from the system of the MC particles. The new EC density is added to the existing one and participates normally in the system \eqref{eq:dcc}--\eqref{eq:ecm}. Conversely, a part of the EC density undergoes EMT towards MC density which is then transformed into particles via a \textit{density-to-particle} operator defined in Section \ref{sec:part-dens-trans}. These newly formed MCs are then added to the system of the existing MC particles.
	
Moreover, at every (time instance and) timestep of the method, the full distribution of MC particles is transformed temporarily to density (without undergoing MET to ECs), via the \textit{particle-to-density} operator. They participate then in the proliferation of the ECs, the production of the MMPs, and the degradation on the ECM, cf. \eqref{eq:dcc}--\eqref{eq:ecm}. We give more details on the combination of the EC and MC phases in Section \ref{sec:hybrid}.
%-----------------------------------------------------------------------------------------------------------------------
\subsection{Modelling phase transitions between particles and densities}\label{sec:part-dens-trans}

In this section we describe the forward \textit{particle-to-density} and the backward \textit{density-to-particle} phase transition operators. 

We assume at first, that the domain $\Omega$ is regular enough to be \textit{uniformly partitioned} in equal rectangles/\textit{partition cells} $\{M_i, i\in I\}$
\begin{equation} \label{eq:partition}
	\Omega = \bigcup_{i\in I} M_i,
\end{equation}
where every $M_i$ is an affine translation of the \textit{generator cell} $K_0$. Note that $K_0$ is the same as the support of the characteristic function in \eqref{eq:zeta}. Clearly $|M_i|=|K_0|= K> 0$. 

\begin{remark}
The partition cells $M_i$, $i\in I$ should not be confused with the discretization cells of the numerical method used to solve \eqref{eq:dcc}--\eqref{eq:ecm}. The latter constitute an instance of a sequence of computational grids of zero-converging step size, whereas the former have a step size that represents physical properties of biological cells\footnote{We mostly consider the diameter of the HeLa cell.} and remain fixed for all computational grid resolutions.
\end{remark}

Using the partitioning of $\Omega$ to $\{M_i,i\in I\}$, we \textit{represent} every measurable $c: \Omega\rightarrow \R$ by its \textit{simple-function} decomposition 
\begin{equation}\label{eq:FVlike}
	\sum_{i\in I} c_i(t) \mathcal X _{M_i}(\vec x),
\end{equation}
where $\mathcal X_{M_i}$ is the characteristic function of the set $M_i\subset\Omega$, and $c_i(t)$ the mean value of $c(\cdot,t)$ over $M_i$
\begin{equation}\label{eq:aver}
	c_i(t)=\frac{1}{K}\int_{M_i} c(\vec x ,t) d\vec x.
\end{equation}
Clearly, this representation conserves the mass of $c(\cdot,t)$
\begin{equation}
	\sum_{i\in I}K c_i (\cdot,t)= \int_\Omega c(\vec x, t) d\vec x
\end{equation}

On the other hand a particle, indexed here by $p\in P$, can be represented either by its position and mass (\textit{particle formulation})
\begin{equation}\label{eq:part}
	\(\vec x_p(t),\, m_p(t)\),
\end{equation}
or by the characteristic function with density value (\textit{density formulation})
\begin{equation}\label{eq:parcel}
	\frac{m_p(t)}{K}\mathcal X_{K_p}(\vec x),\\
\end{equation}
where $K_p$ is the affine translation of the generator cell $K_0$ centred at $\vec x_p$. Clearly \eqref{eq:parcel} implies that the mass $m_p$ of the particle is uniformly distributed over $K_p$.

Although the $K_p$, $p\in P$ and the $M_i$, $i\in I$ in \eqref{eq:FVlike} are equivalent up to affine translations (to the $K_0$), they do not in general coincide. The $M_i$, $i\in I$ form a fixed partition of the domain, cf. \eqref{eq:partition}, whereas the $K_p$, $p\in P$ ``follow'' the position of the particles \eqref{eq:parcel}.

Based on the ``dual'' description \eqref{eq:part} and \eqref{eq:parcel} of the particles, we set forth the \textit{transition operators} between particles and densities. 
%\paragraph{\textbf{From particles to density.}}
\begin{figure}
	\centering
	\includegraphics[width=0.4\linewidth]{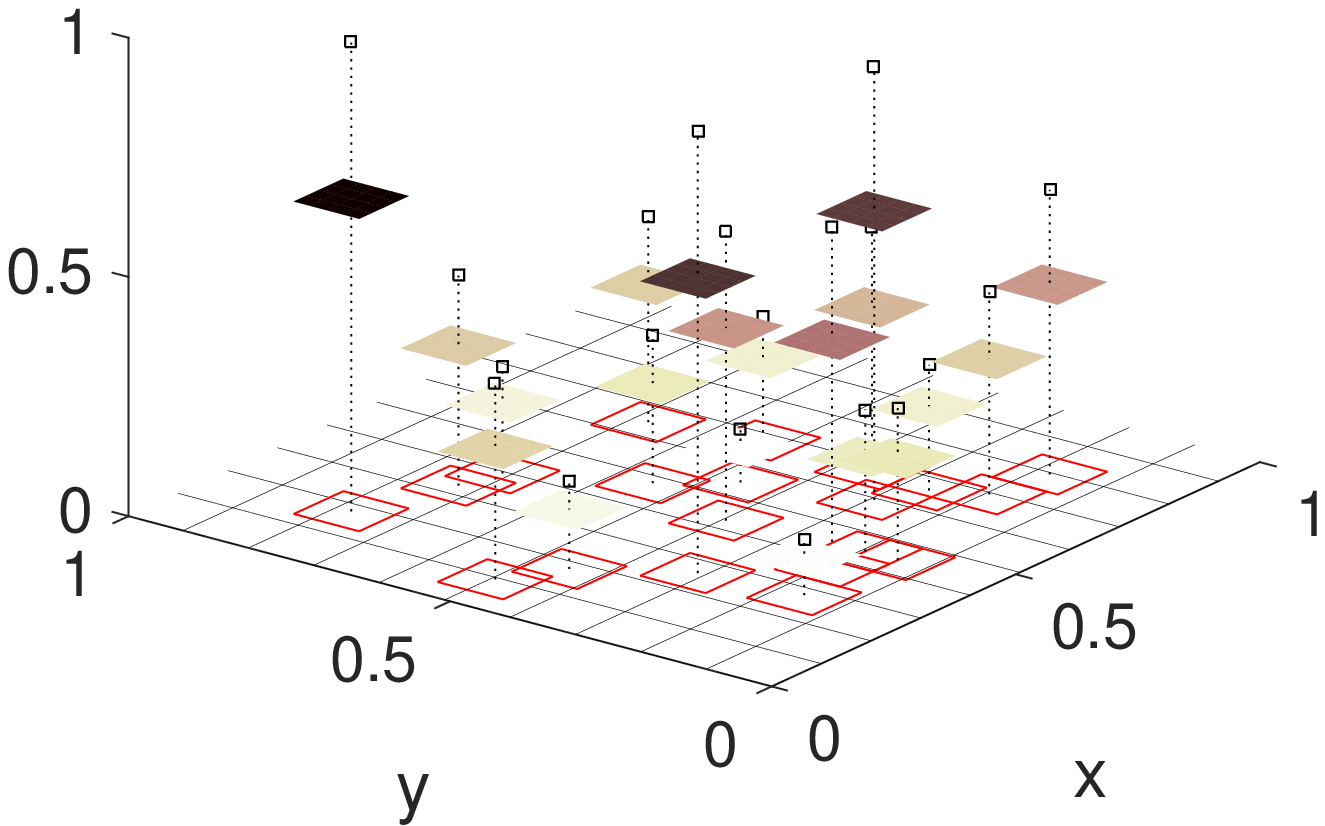} \includegraphics[width=0.4\linewidth]{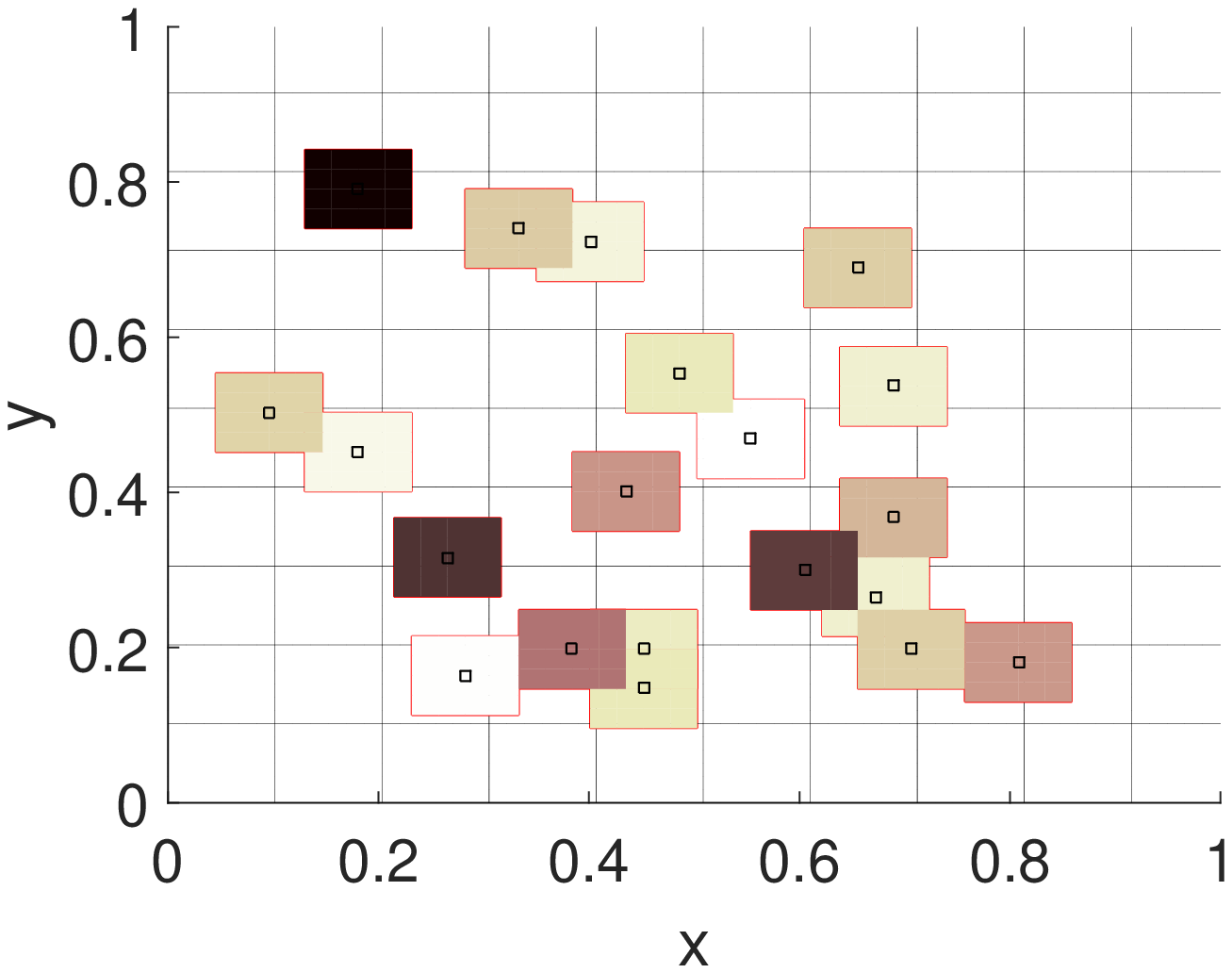}
	\caption{{Two-dimensional graphic} representation of the forward particle-to-density operator $\mathcal F$. (left:) We consider a support $K_p$ ($p\in P$) around the location $x_p$ of every particle. The mass of every particle $m_p$, (shown as points) is uniformly distributed  over the respective support $K_p$. The grid represents the partitioning of the domain. (right:) A view from above reveals that the supports $K_p$ can overlap with several cells of the partition. The corresponding masses are assigned to the partition cells using \eqref{eq:part2dens}.}
	\label{fig:part2dens}
\end{figure}

\subsubsection{Particles to density transition}
Let $\left\{ (\vec x_p(t), m_p(t)), \; p\in P\right\}$ be a collection of particles. Using \eqref{eq:part.better}, we define the \textit{forward particle-to-density operator} $\mathcal F$,
\begin{equation}\label{eq:inter.fwd}
	\left\{ (\vec x_p(t), m_p(t)), \ p\in P\right\} \stackrel{\mathcal F}{\longrightarrow} c(\vec x,t).
\end{equation}
To define the target function $c(\vec x,t)$, we go through all the particles, indexed here by $p\in P$, and consider their corresponding density formulation \eqref{eq:parcel}. The support $K_p$ of the particles, overlaps with (possibly) several\footnote{Since the sets $K_p$, $p\in P$ and $M_i$, $i\in I$ are two-dimensional quadrilaterals of the same dimensions, every $K_p$ overlaps at most four $M_i$s.} of the partition cells $M_i$, $i\in I$. In each of these partition cells, we assign the corresponding portion of the particle mass  
\begin{equation}\label{eq:parcel.decomp}
	m_p\big |_{M_i} = \frac{m_p}{K} \big|K_p\cap M_i\big|.
\end{equation}
Next, we account for the contribution of all particles $p\in P$ at the cell $M_i$ by
\begin{equation}\label{eq:part2dens}
	c_i (t) = \sum_{p\in P} \frac{1}{K}m_p\big |_{M_i}
	\stackrel{\text{\eqref{eq:parcel.decomp}}}{=} \sum_{p\in P} \frac{m_p(t)}{K^2} \big|K_p\cap M_i\big|, \quad \text{ for }i\in I\,.
\end{equation}

In view now of \eqref{eq:FVlike} and \eqref{eq:part2dens}, we deduce the density function $c(\vec x,t)$ (as a simple function) over the full domain $\Omega$ as
\begin{equation}\label{eq:dens.from.part}
	c(\vec x,t) = \sum_{i_\in I}c_i(t) \mathcal X_{M_i}(\vec x), \quad x\in\Omega.
\end{equation}
Refer to Fig. \ref{fig:part2dens} for a graphical representation of the forward particle-to-density operator $\mathcal F$ {in two dimensions}.

\subsubsection{Density to particles transition}
\begin{figure}
	\centering
	\includegraphics[width=0.4\linewidth]{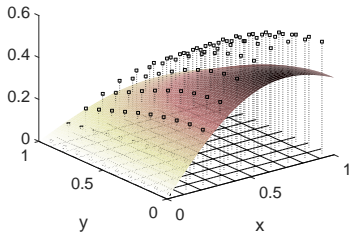}
	\caption{Graphical representation of the backward density-to-particle operator $\mathcal B$. We compute the mass $m_i$ of the density function $c(\vec x, t)$ (surface), over every partition cell $M_i$, $i\in I$ (quadrilateral grid on the $xy$ plane), using \eqref{eq:dens2part.mass}. We then define the particle as $(\vec x_i, m_i)$ where the location $\vec x_i$ is given by \eqref{eq:dens2part.loc}.}
	\label{fig:dens2part}
\end{figure}

Conversely, we define the \textit{backward density-to-particle operator} $\mathcal B$ for a given density function $c(\vec x, t)$ by
\begin{equation}\label{eq:inter.bwd}
	\left\{ (\vec x_p(t), m_p(t)), \ p\in P\right\}\stackrel{\mathcal  B}{\longleftarrow} c(\vec x, t),
\end{equation}
in the following way: in every partition cell $M_i$, $i\in I$, we assign one particle with mass  
\begin{equation}\label{eq:dens2part.mass}
	m_i(t)=\int_{M_i} c(\vec x,t)d\vec x.
\end{equation}
and position 
\begin{equation}\label{eq:dens2part.loc}
	\vec x_i(t) = \text{ the (bary)centre  of } M_i.
\end{equation}
For practical considerations, we set in the numerical simulations a minimum threshold value on the densities, below which no transition to particles takes place. This threshold value is quite small and is used to avoid very large number of particles of negligible-mass. Refer to Fig. \ref{fig:dens2part} for a graphical representation of the backward density-to-particles operator.

%--------------------------------------------------------------------------------------------------------------------------
\subsection{Combination of the two phases}\label{sec:hybrid}
We denote again the two types of cancer cells, EC and MC by the supescripts $\a$ and $\b$ respectively, and consider for $t\geq 0$ the vector formulation \eqref{eq:w.model} of the system \eqref{eq:dcc}--\eqref{eq:ecm} with the density variables 
$$\vec w(\vec x,t)=\(c^\a(\vec x,t), m(\vec x,t), v(\vec x,t)\).$$
%\begin{align}
%	\frac{\pd}{\pd t} c^\a(\vec x, t) =& D_\a\Delta c^\a(\vec x, t) + \rho^\a_c c^\a(\vec x, t)\(1-c^\a(\vec x, t)-c^\b(\vec x, t) - v(\vec x, t)\),
%	\nonumber \\
%	\frac{\pd}{\pd t} m(\vec x, t) =& D_m\Delta m(\vec x, t) +\rho_m^\a c^\a(\vec x, t)+\rho_m^\b c^\b(\vec x, t) - \lambda_m m(\vec x, t),
%	\nonumber \\
%	\frac{\pd}{\pd t} v(\vec x,t) =& - \(\lambda_v^\a c^\a(\vec x,t) + \lambda_v^\b c^\b(\vec x,t)\) m(\vec x, t)v(\vec x,t),
%	\nonumber
%\end{align}
At the same physical time $t$, we write the MC particles as 
\begin{equation}
	\mathcal P^\b(t)=\left\{\(\vec x_p^\b(t),m_p^\b\),\, p\in P(t)\right\},
\end{equation}
and, accordingly, the overall system is given by the tuple 
\begin{equation}\label{eq:tuple}
	\(\vec w(\vec x, t), \mathcal P^\b(t)\),\ \vec x \in \Omega, \ t\geq 0.
\end{equation}
In the evolution of the overall system, we consider the EMT and MET processes separately from the rest of the dynamics of the system \eqref{eq:dcc}--\eqref{eq:ecm}\footnote{To ease the presentation and since the EMT and MET are assumed to {be} \textit{instantaneous} and \textit{tautochronous}, we drop the dependence of the density variables and the particles on $\vec x$ and/or $t$.}.

\subsubsection{EMT operator}
The EMT triggering mechanism is not one of the main foci of this work. Instead, we assume a simplified approach where a randomly chosen part of the ECs (in density formulation) $c^\a_\text{EMT}$ undergoes EMT to give rise to MCs (still in density formulation)
$$ c^\a_\text{EMT} \xrightarrow{\text{EMT}}c^\b_\text{EMT}.$$
The newly created  MC density $c^\b_\text{EMT}$ is transformed to MC particles via the density-to-particle operator $\mathcal B$ given in \eqref{eq:inter.bwd}
\begin{equation}
	c^\b_\text{EMT} \stackrel{\mathcal B}{\longrightarrow} \left\{ (\vec x_p^\b,  m_p^\b),\ p\in P^\text{EMT}\right\},
\end{equation}
where $\vec x_p^\b$, $m_p^\b$ follow from \eqref{eq:dens2part.mass}, \eqref{eq:dens2part.loc} and $P^\text{EMT}$ is the corresponding set of indexes. Subsequently, the family of MC particles is updated as the amalgam of the existing and the newly created particles
\begin{equation}
\underbrace{\left\{ (\vec x_p^\b,m_p^\b),\ p\in P\right\}}_\subtext{existing MC particles} \uplus 
\underbrace{\left\{ (\vec x_p^\b,m_p^\b),\ p \in P^\text{EMT} \right\}}_\subtext{newly created MC particles} =\left\{ (\vec x_p^\b,m_p^\b),\ p\in P^\text{new} \right\},
\end{equation}
where $P^\text{new}$ is a re-enumeration of the multiset $P \uplus P^\text{EMT}$.

Overall, combining the density and particle phases, the EMT operator reads
\begin{equation}\label{eq:oper.EMT}
%\mathcal R^\text{EMT}\( c^\a (\vec x,t), \left\{\(\vec x_p^\b(t),m_p^\b\),\ p\in P(t)\right\}\) =
%\( c^\a(\vec x,t) - c^\a_\text{EMT}(\vec x,t) , \left\{ (\vec x_p^\b(t),m_p^\b),\ p\in P^\text{new}(t) \right\} \).
\mathcal R^\text{EMT}\( c^\a , \left\{\(\vec x_p^\b,m_p^\b\),\ p\in P\right\}\) =
\( c^\a - c^\a_\text{EMT} , \left\{ (\vec x_p^\b,m_p^\b),\ p\in P^\text{new}\right\} \).
\end{equation}

%-----------------------------------------------------------
\subsubsection{MET operator}
	As with the EMT, the triggering mechanism of the MET is not one of the foci of this paper. We instead assume an approach where each of the MC particles $\left\{ \(\vec x_p^\b,  m_p^\b\),\ p\in P\right\}$ undergoes MET to ECs randomly
\begin{equation}
	\left\{ (\vec x_p^\b,  m_p^\b),\ p\in P\right\} \xrightarrow{\text{MET}} \underbrace{\left\{ (\vec x_p^\a,  m_p^\a),\ p\in P^\text{MET} \right\}}_\subtext{newly created EC particles}.
\end{equation}
The resulting EC particles are instantaneously transformed to density via the particle-to-density operator $\mathcal F$ given in \eqref{eq:inter.fwd}:
$$\left\{ (\vec x_p^\a,  m_p^\a),\ p\in P^\text{MET} \right\} \stackrel{\mathcal F}{\longrightarrow}  c^\a_\text{MET}.$$
In operator form, the MET reads
\begin{equation}\label{eq:oper.MET}
\mathcal R^\text{MET}\( c^\a , \left\{\(\vec x_p^\b,m_p^\b\),\ p\in P\right\}\) =
\( c^\a + c^\a_\text{MET} , \left\{ (\vec x_p^\b,m_p^\b),\ p\in P^\text{new}\right\} \),
\end{equation}
where $P^\text{new}$ is a re-enumeration of the set difference $P \setminus P^\text{MET}$.

\subsection{Combination of the two phases}
The evolution of the overall system of ECs is controlled by \eqref{eq:dcc}--\eqref{eq:ecm} and \eqref{eq:brownian.1}. We study this combined system of PDEs and SDEs numerically and postpone any analytical investigations to a follow up work.  To this end, we consider the model \eqref{eq:w.model} and first set 
\begin{align*}
	\vec W^n =& \left\{\vec w_{(i,j)}^n=\(c_{(i,j)}^n, m_{(i,j)}^n, v_{(i,j)}^n\),\ (i,j)\in M_x \times M_y\right\},\\
	\mathcal P^{\b,n}=&\left\{\(\vec x_p^{\b,n},m_p^\b\),\ p\in P^n\right\},
\end{align*}
to denote numerical approximations of the density and particle variables $\vec w(\vec x, t)$ and $\mathcal P^\b(t)$, respectively, at the {instantaneous time} $t=t^n$. Here, $M_x$ and $M_y$ denote the resolution of the grid along the $xy-$ directions. We refer to Appendix \ref{sec:FV} for further information on the numerical method employed on  $\vec W$; we focus here on the combination of the two phases by considering an \textit{operator splitting} approach. In particular, for $t\in[t^n, t^{n+1}]$, $t^{n+1}=t^n+\tau^n$, we assume that:
\begin{itemize}
\item[---]
	During the time period $[t^n, t^n + \frac{1}{2}\tau^n]$, the system evolves, \textit{without} the influence of the EMT or the MET, as
	\begin{subequations}
	\begin{equation}
		\(\vec W^n, \mathcal P^{\b,n} \)\longrightarrow \(\vec W^{n+1/2}, \mathcal P^{\b,n+1/2}\)
	\end{equation}
	with
	\begin{align}
		\vec W^{n+1/2} = &\mathcal N^{\left[t^n,t^n+\frac{1}{2}\tau^n \right]} \(\vec W^{n}, \mathcal P^{\b,n}\),\\
		\mathcal P^{\b,n+1/2} = &\left\{\(\vec x_p^{\b,n+1/2},m_p^{\b,n+1/2}\),\ p\in P^{n+1/2}\right\},
	\end{align}
	\end{subequations}
	{where} $\mathcal N^{[t,t+\tau]}$ is the numerical solution operator responsible for the spatiotemporal evolution of the system \eqref{eq:dcc}--\eqref{eq:ecm} ---without EMT and MET. {Here}, the $x_p^{\b,n+1/2}$, $p\in P^{n}$ is given by the It\^o-type particle motion scheme \eqref{eq:part.loc}, re-written here with respect to the local variables
	\begin{equation}
		\vec x_p^{\b,n+1/2} = \vec x_p^{\b,n} + \vec A\(\vec x_p^{\b,n}\) \frac{1}{2} \tau^n + \vec B\(\vec x_p^{\b,n}\)\cdot \vec Z_p \sqrt{\frac{1}{2} \tau^n}.
	\end{equation}
	The number of particles, their {indices} and masses remain unchanged during this step $[t^n, t^n + \frac{1}{2} \tau^n ]$, i.e. 
	$$P^{n+1/2}=P^{n}\quad \text{and}\quad m_p^{\b,n+1/2}= m_p^{\b,n},\quad \forall p\in P^{n}.$$	
	Altogether, the combined evolution operators of the two phases reads for this time period as:
	\begin{equation}\label{eq:dscr.str.l}
		\mathcal M_{\frac{1}{2} \tau^n}\(\vec W^n,\, \mathcal P^{\b,n} \) = \(\vec W^{n+1/2},\, \mathcal P^{\b,n+1/2}\).
	\end{equation}
		
\item[---]
	At $t = t^n + \frac{1}{2} \tau^n$, the EMT and MET processes take place; they are assumed to be \textit{instantaneous} and \textit{tautochrone}. They are represented by the $\mathcal R^\text{EMT}$ and $\mathcal R^\text{MET}$ operators introduced in \eqref{eq:oper.EMT} and \eqref{eq:oper.MET} respectively. For consistency, we scale them by the time step $\tau^n$ and change their notation to $\mathcal R^\text{EMT}_{\tau^n}$ and $\mathcal R^\text{MET}_{\tau^n}$, respectively. 
	
	In effect, the tuple $\(\vec W^{n+1/2},\, \mathcal P^{\b, n+1/2}\)$ develops as
	\begin{equation}\label{eq:dscr.str.c}
		\(\tilde{\vec W}^{n+1/2},\,\tilde{\mathcal P}^{\b, n+1/2}\) = \mathcal R_{\tau^n} \(\vec W^{n+1/2},\,\mathcal P^{\b, n+1/2}\),
	\end{equation}
	where $\mathcal R_{\tau^n}$ denotes the parallel application of $\mathcal R^\text{EMT}_{\tau^n}$ and $\mathcal R^\text{MET}_{\tau^n}$\footnote{Note that the $\mathcal R^\text{EMT}_{\tau^n}$ acts on the EC density and the $\mathcal R^\text{MET}_{\tau^n}$ on the MC particles}.
		
\item[---]
	During $[t^n + \frac{1}{2} \tau^n, t^{n+1}]$, the two phases evolve again without the influence of the EMT and MET as
	$$\(\tilde{\vec W}^{n+1/2},\,\tilde{\mathcal P}^{\b, n+1/2}\) \longrightarrow \(\vec W^{n+1},\, \mathcal P^{\b,n+1}\),$$
	where, in a similar way as in $[t^n, t^n + \frac{1}{2} \tau^n]$,
	\begin{align}
		\vec W^{n+1}=&\mathcal N^{\left[t^n+\frac{1}{2} \tau^n, t^{n+1}\right]} \(\vec W^{n+1/2},\, \tilde{\mathcal P}^{\b, n+1/2}\),\\
		\mathcal P^{\b,n+1}=&\left\{\(\vec x_p^{\b,n+1},m_p^{\b,n+1}\),\ p\in P^{n+1}\right\}.
	\end{align}
	Again, $\mathcal N^{\left[t^n+\frac{1}{2} \tau^n, t^{n+1}\right]}$ represents the numerical method for the solution of the system \eqref{eq:dcc}--\eqref{eq:ecm}, 
	$$P^{n+1}=P^{n+1/2}\quad \text{and}\quad m_p^{\b,n+1}= m_p^{\b,n+1/2},\quad \forall p\in P^{n+1}.$$
	In the above, $\vec x_p^{\b,n+1}$, $p\in P^{n+1/2}$, is given by the Ito-type scheme \eqref{eq:part.loc}
	\begin{equation}
		\vec x_p^{\b,n+1} = \vec x_p^{\b,n+1/2} + \vec A\(\vec x_p^{\b,n+1/2}\) \frac{\tau^n}{2} + \vec B\(\vec x_p^{\b,n+1/2}\)\cdot \vec Z_p \sqrt{\frac{\tau^n}{2}}.
	\end{equation}	
	We combine the evolution operators of the two phases as:
	\begin{equation}\label{eq:dscr.str.r}
		\mathcal M_{\frac{\tau}{2}}\(\tilde{\vec W}^{n+1/2},\,\tilde{\mathcal P}^{\b, n+1/2}\) = \(\vec W^{n+1},\, \mathcal P^{\b,n+1}\).
	\end{equation}
\end{itemize}

Overall, using \eqref{eq:dscr.str.l}, \eqref{eq:dscr.str.c}, and \eqref{eq:dscr.str.r}, we can write the combined evolution operator for the time period $[t^n, t^{n+1}]$ as a splitting method of the form
\begin{equation}\label{eq:oper.model}
	\(\vec W^{n+1},\, \mathcal P^{\b,n+1}\) 
		= \mathcal M_{\frac{\tau^n}{2}} \; \mathcal R_{\tau^n}\; \mathcal M_{\frac{\tau^n}{2}} \(\vec W^{n},\, \mathcal P^{\b,n}\).
\end{equation}

%------------------------------------------------------------------------------------------------------
\section{Experiments and simulations}\label{sec:simulations}
We perform and present three numerical experiments to exhibit the dynamics and combination of the two phases. As the focus of this paper is more technical, we postpone the more biological relevant experiments and discussion for a follow-up work. 

The implementations of numerical schemes and algorithms, and the simulations of the experiments included in this paper have been conducted in \cite{MATLAB:2015b}.
%--------------------------
\begin{experiment}[EMT and particle flow]\label{exp:exp.1} \normalfont
	\begin{table}[t]%[tbhp]
	\caption{Parameters and units corresponding to Experiment \ref{exp:exp.1} and Fig. \ref{fig:exp.1}.}
	\label{tbl:exp.1}
	\centering
	\scriptsize
	\begin{tabular}{l|c|l}
		description&symbol&values and units\\
		\hline
		EC dens. diff.	coef.	&$D_\a$				&$0$ $\mathrm{cm^2d^{-1}}$			
		
		\\
		EC dens. prol.	coef.	&$\rho^\a$			&$0$ $\mathrm{d^{-1}}$
		\\
		MC part. diff. 	coef.	&$|\vec B|$			&$1.6$ $\mathrm{cm^2d^{-1}}$
		\\
		MC part. hapt. 	coef.	&					&$30$ $\mathrm{cm^3 mol^{-1}d^{-1}}$				
		\\
		MC part. ref. mass		&$m_\text{ref}$		&$1\times 10^{-5}$ $\mathrm{gr}$
		\\
		MC part. ref. diam.		&$|K_0|$			&$1\times 10^{-2}$ $\mathrm{cm}$
		\\
		EMT prob.				&					&$5\times 10^{-4}$ 
		\\
		EMT rate				&$\mu_a$			&$1\times 10^3$
		\\
		MET prob.				&					&$0$
		\\
		MMP diff. coef.	 		&$D_m$				&$0$
		\\
		ECM EC dens. degr.		&$\lambda_v^\a$		&$0$ $\mathrm{cm^2mol^{-1}d^{-1}}$
		\\
		ECM MC dens. degr.		&$\lambda_v^\b$		&$0$ $\mathrm{cm^2mol^{-1}d^{-1}}$
	\end{tabular}
	\end{table}	
	We set $\Omega = [-2,2]\times[-2,2]$ and consider the initial EC density
	\begin{subequations}
	\begin{equation}
		c^\a(\vec x,0) = \(e^{-5(x_1^2+x_2^2)}-0.7\)^+\,,
	\end{equation}
	with $\vec x=(x_1,x_2)\in\Omega$, where $\(\cdot\)^+$ denotes the positive part function. The ECM is non-uniform and exhibits a gradient towards the upper-right part of the domain
	\begin{equation}
	v(\vec x,0) =  0.045\(2x_1+3x_2\) + 0.45.
	\end{equation}
	\end{subequations}
	Initially, no MC particles nor MMPs are present. 

	We close the system with \textit{no-flux boundary conditions} for the EC and MMP density and \textit{reflective boundary conditions}\footnote{Each particle that escapes the domain is returned to its last position within the domain. It then ``decides'' its new direction randomly.} for the MCs particles. As the ECM is modelled as an immovable part of the system, it does not translocate, hence no boundary conditions are needed. The parameters for this experiment are given in Table \ref{tbl:exp.1} and the result of the simulation is shown in Fig. \ref{fig:exp.1}.

	The phenomena that is observed in this experiment can be described as follows:

	The ECs undergo EMT to MC and new particles appear in the system. The particles ``sense'' the gradient of the ECM and respond haptotactically to it. Their motion incorporates also a random component; the resulting migration is a biased-random motion. The simplified EMT that we assume in this experiment, takes place in every partition cell $K_i$, $i\in I$ with a probability that is denoted as ``EMT prob.'' in Table \ref{tbl:exp.1}. The set union of all the partitions cells where EMT takes place defines the set $\mathcal A$ that appears in \eqref{eq:dcc}. The rate $\mu_a$ at which the EMT occurs in $\mathcal A$ is a given constant. There is no proliferation of the ECs assumed in this experiment, neither diffusion nor MET. Hence, the losses that the ECs suffer due to the EMT appear as ``holes'' in their density profile and are not replenished with time.

	\begin{figure}
		\centering
		\subfigure[$t=0.01$]{\includegraphics[width=0.28\linewidth]{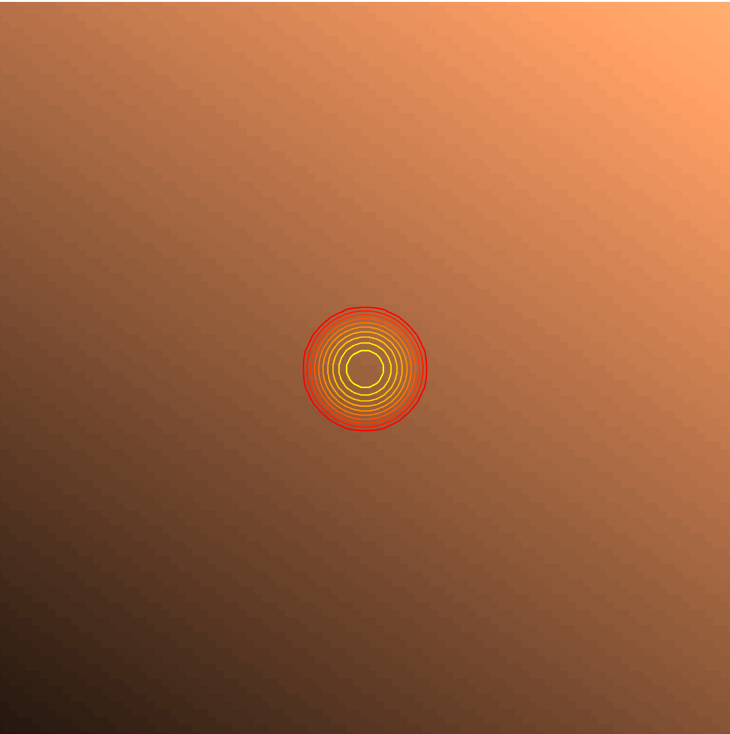}}
		\subfigure[$t=0.06$]{\includegraphics[width=0.28\linewidth]{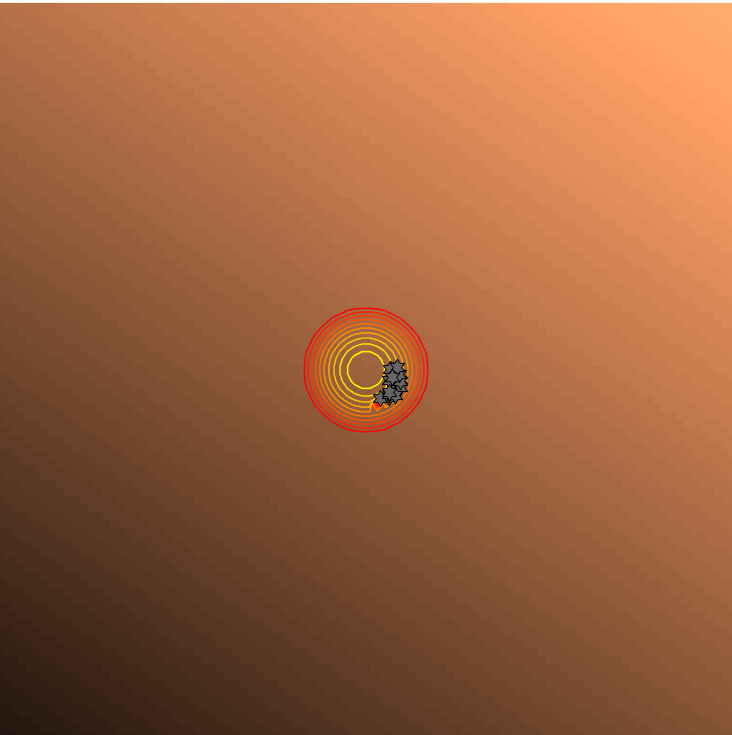}}
		\subfigure[$t=0.10$]{\includegraphics[width=0.28\linewidth]{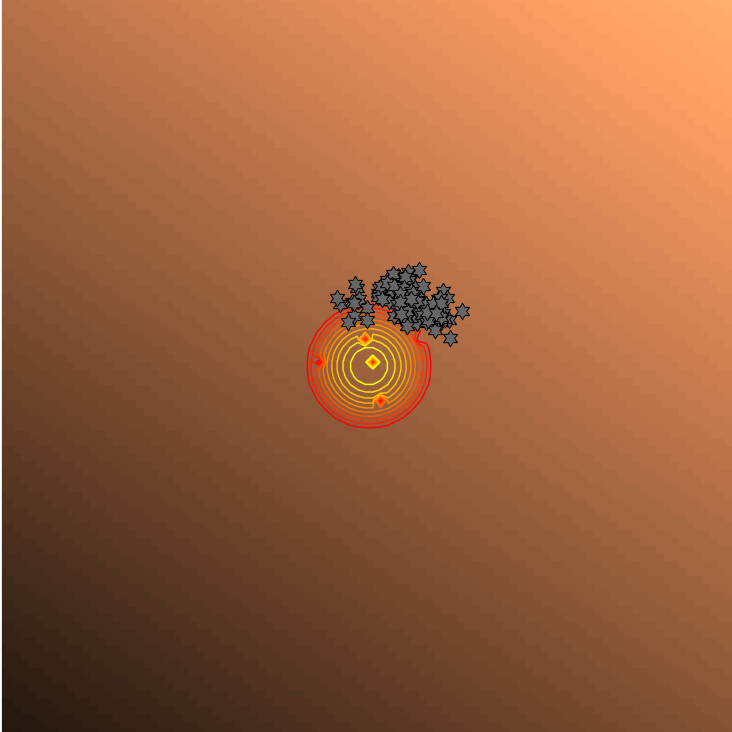}}\\
		\subfigure[$t=0.40$]{\includegraphics[width=0.28\linewidth]{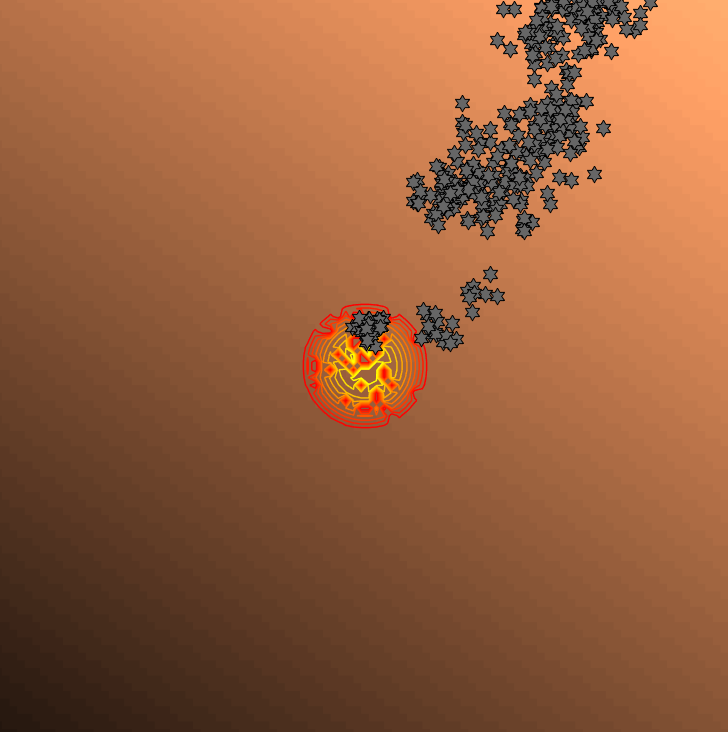}}
		\subfigure[$t=0.85$]{\includegraphics[width=0.28\linewidth]{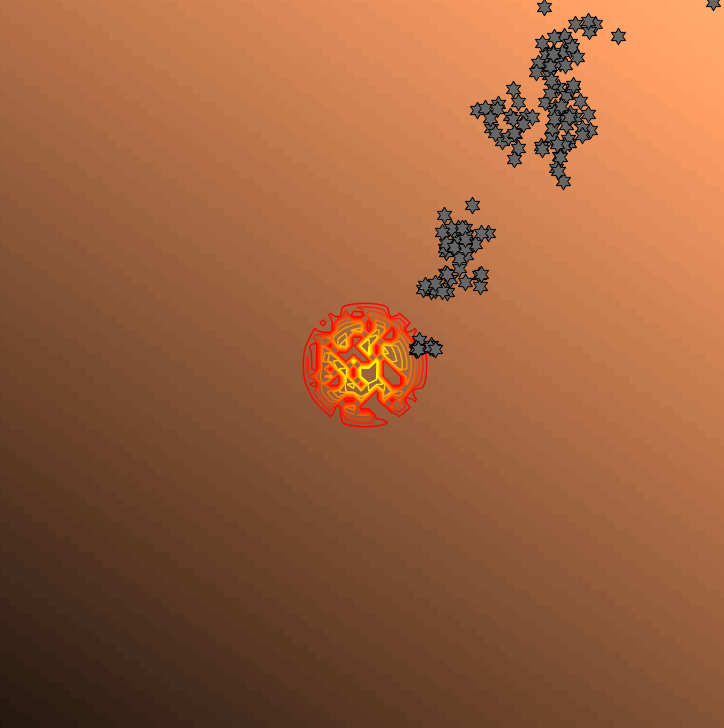}}~~
		\subfigure[colorbars]{\raisebox{-0.25em}{\includegraphics[height=10.75em]{{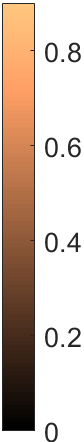}}}
			\includegraphics[height=10.45em]{{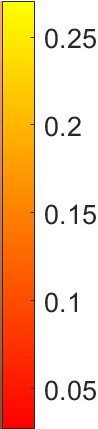}}}
		\caption{Experiment \ref{exp:exp.1} (EMT and flow). The experiment is visualised over $[-1,1]^2$. (a): An initial circular EC tumour (shown by isolines) resides over an ECM that exhibits a gradient towards the north-east direction. (b): The DDC density undergoes EMT and gives rise to clusters of MC particles (black stars). (c): Due to the diffusion and the haptotaxis, the particles escape the initial tumour and migrate along the gradient of the ECM. (d): We do not assume proliferation for the ECs, hence the losses of their densities that suffer towards MCs are not replenished. This gives rise to ``holes'' in the initial tumour. (e): The phenomenon continues as long as parts of the EC density transform to MC particles. (f): Common colorbars for the ECM (left) and the EC densities (right) in all sub-figures.}\label{fig:exp.1}
	\end{figure}

\end{experiment}

%---------------
\begin{experiment}[Self-generated gradient]\label{exp:exp.8}\normalfont
		\begin{table}[t]%[tbhp]
		\caption{Parameters and units corresponding to Experiment \ref{exp:exp.8} and Fig. \ref{fig:exp.8}.}
		\label{tbl:exp_8}
		\centering
		\scriptsize
		\begin{tabular}{l|c|l}
			description&symbol&values and units\\
			\hline
			EC dens. diff.	coef.	&$D_\a$				&$0$ $\mathrm{cm^2d^{-1}}$					
			\\
			EC dens. prol.	coef.	&$\rho^\a$			&$0$ $\mathrm{d^{-1}}$				
			\\
			MC part. diff. 	coef.	&$|\vec B|$			&$2\times 10^{-2}$ $\mathrm{cm^2d^{-1}}$			
			\\
			MC part. hapt. 	coef.	&					&$1\times 10^{-3}$ $\mathrm{cm^3mol^{-1}d^{-1}}$
			\\
			MC part. ref. mass		&$m_\text{ref}$		&$3\times 10^{-9}$ $\mathrm gr$
			\\
			MC part. ref. diam.		&$|K_0|$			&$1\times 10^{-3}$ $\mathrm cm$
			\\
			EMT prob.			&					&$1$ 
			\\
			EMT rate			&$\mu_a$			&$10$
			\\
			MET prob.			&					&$0$
			\\
			MMP diff. coef.	&$D_m$				&$0$
			\\
			ECM EC dens. degr.	&$\lambda_v^\a$		&$20$ $\mathrm{cm^2mol^{-1}d^{-1}}$
			\\
			ECM MC dens. degr.	&$\lambda_v^\b$		&$200$ $\mathrm{cm^2mol^{-1}d^{-1}}$
		\end{tabular}	
	\end{table}	
	A typical phenomenon that macroscopic cancer invasion models exhibit, is the appearance of a propagating front that invades the ECM faster than the rest of the tumour, see e.g \cite{Chaplain.1999,Chaplain.2005,Sfakianakis.2016b}. This front is followed by an intermediate distribution, whereas the bulk of the tumour stays further behind. This phenomenon is due primarily to the degradation of the extracellular chemical or the ECM landscape by the cancer cells. Such phenomena have been observed previously both in mathematical models and in biological experiments, see for example \cite{Insall.2016, Chaplain.2000}.
	\begin{figure}
	\centering
	\begin{tabular}{ccc}
		\subfigure[$t=1$]{\includegraphics[width=0.25\linewidth]{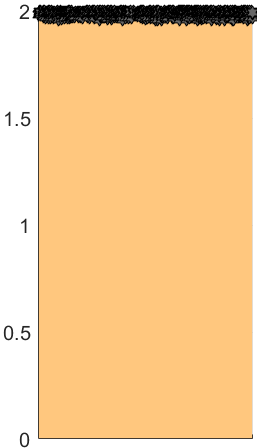}}&
		\subfigure[$t=40$]{\includegraphics[width=0.25\linewidth]{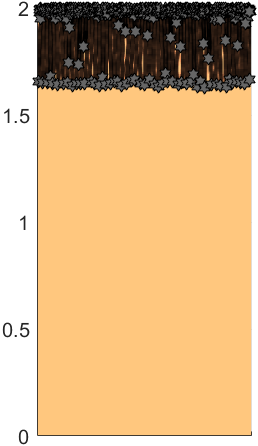}}&
		\subfigure[$t=90$]{\includegraphics[width=0.25\linewidth]{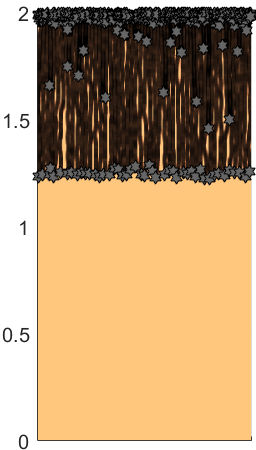}}\\
		\subfigure[$t=130$]{\includegraphics[width=0.25\linewidth]{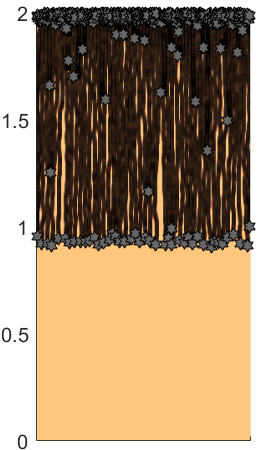}}&
		\subfigure[$t=180$]{\includegraphics[width=0.25\linewidth]{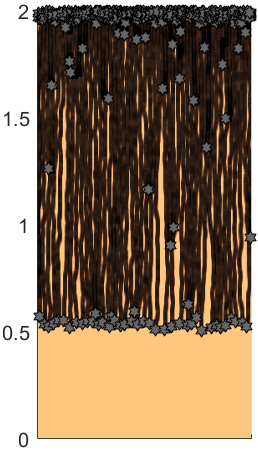}}&
		\subfigure[$t=230$]{\includegraphics[width=0.252\linewidth]{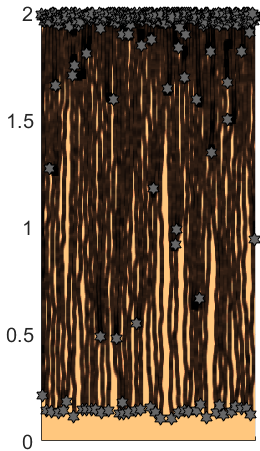}}\\
	\end{tabular}
	\caption{Experiment \ref{exp:exp.8} (Self-generated gradient). (a): Over an initially uniform ECM (background) resides a number of MC particles. As the matrix is initially uniform, the motion of the particles is mostly Brownian. (b): The particles degrade the matrix and create a gradient. This self-induced gradient is sensed by the particles that are closer to the ``interface''. In effect, their motion is mostly haptotaxis driven. (c): As the particles invade the ECM, they continue to produce MMPs, and to degrade the matrix. They  follow the new gradient that they have induced. (d-f): The migration of the particles in the front is persistent in direction and speed, while the particles in the rear (where the ECM is depleted) perform mostly a Brownian motion.}\label{fig:exp.8}
	\end{figure}
	In accordance to these experimental and modelling findings, we exhibit here the ability of the particle description of our model to reproduce such phenomena. In particular, we show that as the cancer cells degrade the ECM, they induce a gradient on it, and subsequently they respond to this gradient by performing a directed and sustainable invasion.
	
	For this experiment we consider the domain $\Omega=[-0.5,0.5]\times[0,2]$, over which lies a uniform ECM 
	\begin{equation}
		v(\vec x,0) =  0.1,\quad \vec x \in \Omega.
	\end{equation}

	On the upper part of the domain, an initial EC density is found
	\begin{equation}
		c^\a(\vec x,0) = 10^{-4}\mathcal X_{S_1}(\vec x)\,,\quad \vec x \in \Omega,
	\end{equation}	
	with $S_1=\left\{\vec x= (x_1,x_2)\in \Omega \, \big|\,  x_2> 0.01\sin(5\pi x_1) + 1.97\right\}$. Before the beginning of the simulation, the EC density $c^\a(\vec x,0)$ is completely transformed to MC particles. The MCs secrete MMPs that participate in the degradation of the ECM. No MET takes place in this experiment. The corresponding modelling parameters are given in Table \ref{tbl:exp_7} and the simulation results in Fig. \ref{fig:exp.8}.

	In view of \eqref{eq:part.loc}, all the particles perform a biased random motion; since the ECM is uniform, this motion is initially purely Brownian. As the ECM is degraded, a gradient in the matrix is formed. The particles that reside closer to this ``interface'' sense the gradient and respond haptotactically to it. The directed part of their motion dominates and drives the particles to higher matrix densities. As the particles continue their invasion of the ECM they keep on producing MMPs, degrading the ECM, and following the newly created gradient. Their motion is persistent in direction and speed.

	We can now address with our model, particular questions of experimental interest: what is the minimum number of cancer cells needed to induce and sustain a consistent in direction and speed invasion of the ECM? How does the remodelling of the matrix affect the self-generated gradient motion? Such questions would among others serve as a bridge between experimental observations and mathematical models. Their study has to be the topic of a follow-up work the relevant experimental data should also be analysed, as was done, for example, in \cite{Madzvamuse.2016}.
	
\end{experiment}

%---------------
\begin{experiment}[ECM invasion]\label{exp:exp.7}\normalfont
	\begin{figure}
		\centering
		{\includegraphics[height=9em]{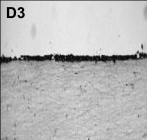}}
		{\includegraphics[height=9em]{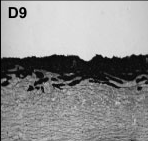}}
		{\includegraphics[height=9em]{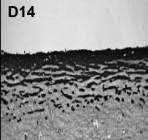}}
		\caption{Timecourse (days 3, 9, and 14) study of the invasion of squamus cell carcinoma cells (black matter) on an organotypic assay with human fibroblast cells (gray matter). The invasion occurs in the form of cancer cell ``islands'' formed in front of the main body of the tumour. We reproduce the same phenomenon in the invasion Experiment \ref{exp:exp.7} and in Fig. \ref{fig:exp.7}.  These images are taken from \cite{Nystroem.2005} \redd{(PERMISSION REQUESTED)}.}\label{fig:Nystrom}
	\end{figure}
	\begin{table}[t]%[tbhp]
	\caption{Parameters, units, and sourses corresponding to Experiment \ref{exp:exp.7} and Fig. \ref{fig:exp.7}.}
	\label{tbl:exp_7}
	\centering
	\scriptsize
	\begin{tabular}{l|c|l||p{13em}}
		description&symbol&values and units&sources\\
		\hline 
		EC dens. diff.	coef.	&$D_\a$				&$8.64\times 10^{-6}$ $\mathrm{cm^2d^{-1}}$   &\cite{Chaplain.2005}				
		\\
		EC dens. prol.	coef.	&$\rho^\a$			&$1.2$ $\mathrm{d^{-1}}$ 	&\cite{Chaplain.1997}
		\\
		MC part. diff. 	coef.	&$|\vec B|$			&$3\times 10^{-1}$ $\mathrm{cm^2d^{-1}}$	&\cite{Stokes.1991}
		\\
		MC part. hapt. 	coef.	&					&$3$ $\mathrm{cm^3mol^{-1}d^{-1}}$	&(our estimate)
		\\
		MC part. ref. mass		&$m_\text{ref}$		&$3\times 10^{-9}$ $\mathrm{gr}$ &B10NUMB3R5 (HeLa cell)
		%		\\ &&& bionumbers.hms.harvard.edu/
		\\
		MC part. ref. diam.		&$|K_0|$			&$1\times 10^{-3}$ $\mathrm{cm}$ &B10NUMB3R5 (HeLa cell)
		\\
		EMT prob.			&					&$1\times 10^{-5}$ &(our estimate)
		\\
		EMT rate			&$\mu_a$			&$4\times 10^{-3}$ &(our estimate)
		\\
		MET prob.			&					&$2\times10{^-2}$ &(our estimate)
		\\
		MMP diff. coef.		&$D_m$				&$0$ %$\mathrm{cm^2d^{-1}}$&
		\\
		ECM EC dens. degr.	&$\lambda_v^\a$		&$1\times10^{-5}$ $\mathrm{cm^2mol^{-1}d^{-1}}$ & \cite{Chaplain.1998}
		\\
		ECM MC dens. degr.	&$\lambda_v^\b$		&$1\times10^{-4}$ $\mathrm{cm^2mol^{-1}d^{-1}}$ & \cite{Chaplain.1998} \& (our estimate)
	\end{tabular}
	\end{table}
	This experiment is motivated by the \textit{organotypic invasion assays} where cancer cells are plated over a  collagen gel that contains healthy tissue, and where their invasion is studied  over time, see for example \cite{Nystroem.2005, Valster.2005} and Fig. \ref{fig:Nystrom}.
	
	We employ the complete set of dynamics of the system and consider the domain $\Omega=[-2,2]^2$ occupied by an ECM of initial density $v(\vec x,0)$ constructed by 64 randomly chosen extremal values per direction that are interpolated in a piecewise linear way. Small perturbations of the form of additive Gaussian noise are also included.

	An initial density of ECs is found in the upper part of the domain
	\begin{equation} 
		c^\a(\vec x,0) = 0.05\,\mathcal X_{S_2}(\vec x)\,,\quad \vec x\in \Omega,
	\end{equation}	
	with $S_2(\vec x)=\left\{\vec x=(x_1,x_2)\in \Omega \, \big|\,  x_2> 0.05\sin(5\pi x_1) + 0.05\,x_1 + 1.1\right\}$.	Initially, neither MC particles nor MMPs exist in the system. The parameters for this experiment can be found in Table \ref{tbl:exp_7} and the simulation results are presented in Fig. \ref{fig:exp.7}.

	The ECs proliferate and diffuse, but most notably transform via EMT to MC particles. These MC particles do not proliferate but they are very aggressive in their motility. As they escape the main body of the tumour, they undergo MET back to ECs. New EC concentrations appear, they grow due to proliferation, and give rise to tumour ``islands''. These ``islands'' merge with each other as well as with the main body of the tumour. The main characteristic and novelty of our hybrid model, is that it predicts, in a natural way, the appearance of these tumour ``islands'' outside of the main body of the tumour.  
	
	The growth of the tumour with the combined dynamics of the ECs and MCs possesses several interesting properties. The tumour grows much faster than it would, if it was comprised only of the ECs. This is so, since the new EC ``islands'' that arise after the MCs have escaped the main body of the tumour, undergo MET, exploit uninhabited locations, and grow ``to all the directions''. On the contrary, in the main body of the tumour, only the ECs found in the periphery contribute to the growth of its support. 
	
	Moreover, the independent and aggressive migration of the MCs provides them with faster access to the circulatory network and the possibility to translocate to secondary places within the organism. As they possess the ability to give rise to EC ``islands'' at the new locations, new tumours might appear, and \textit{metastasis} will have occurred. Although it is not our aim in the current paper to reproduce  particular experimental scenarios, a direct comparison of  the simulation results  in Fig. \ref{fig:exp.7} with the organotypic assay images in Fig. \ref{fig:Nystrom} exhibits clearly that this phenomenon is reproduced in a very natural way by our modelling approach.
	
	Another sought-for property, in cancer invasion modelling, is that the MCs  remain undetected while they invade the ECM. It is not until a new ECs tumour has been established that it can grow to be of any detectable size. Again, this property is reproduced by our modelling approach in a very natural way.
	
	\begin{figure}
	\centering
%	\subfigure[$t=1$]{\includegraphics[height=9em]{Nystrom_03.png}}
%	\subfigure[$t=150$]{\includegraphics[height=9em]{Nystrom_09.png}}
%	\subfigure[$t=200$]{\includegraphics[height=9em]{Nystrom_14.png}}
%	\\
	\subfigure[$t=1$]{\includegraphics[width=0.30\linewidth]{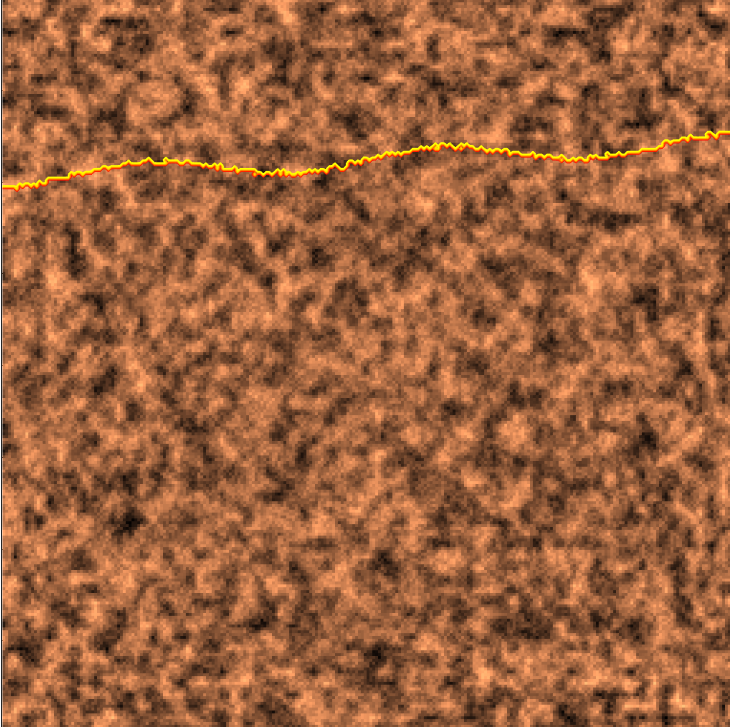}}
	%		\subfigure[$t=100$]{\includegraphics[width=0.28\linewidth]{exp_7_100}}
	\subfigure[$t=150$]{\includegraphics[width=0.30\linewidth]{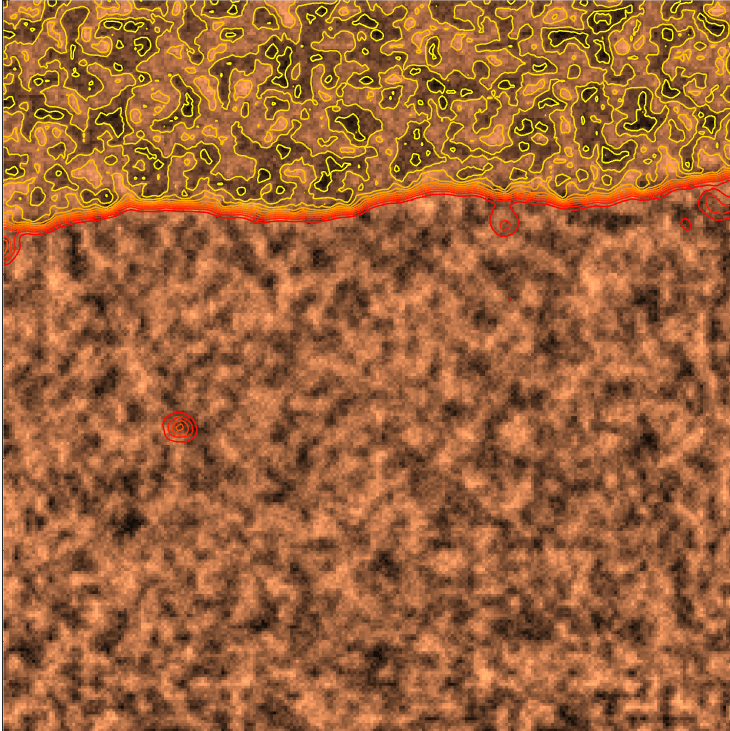}}
	\subfigure[$t=200$]{\includegraphics[width=0.30\linewidth]{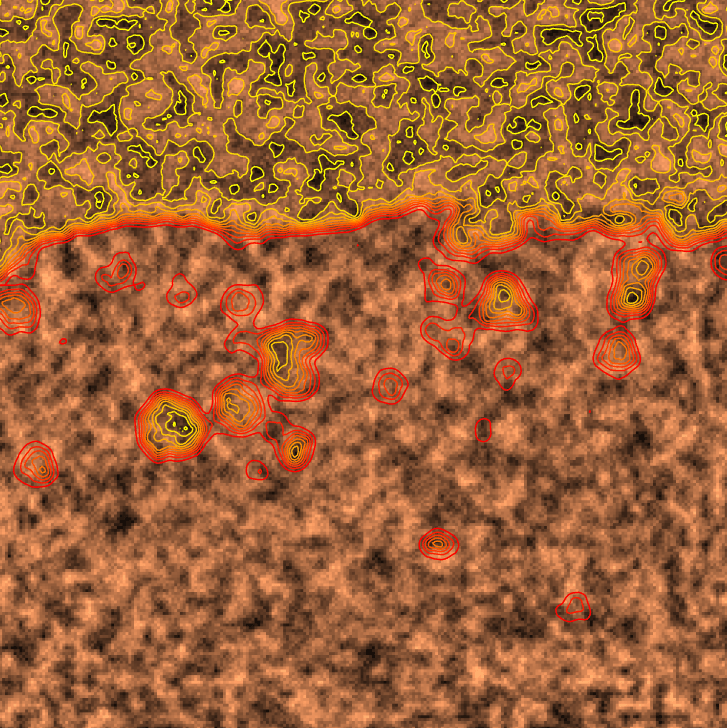}}\\
	\subfigure[$t=270$]{\includegraphics[width=0.30\linewidth]{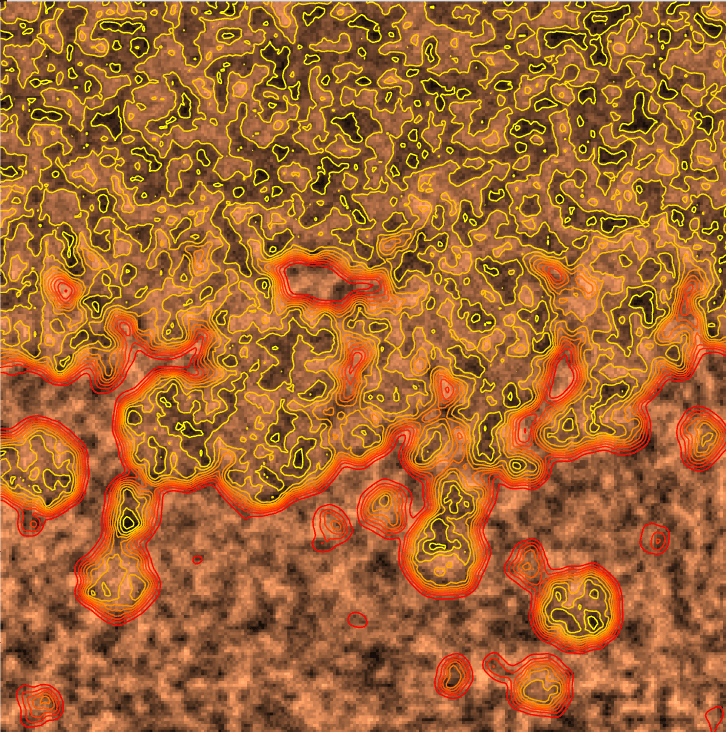}}
	\subfigure[$t=330$]{\includegraphics[width=0.30\linewidth]{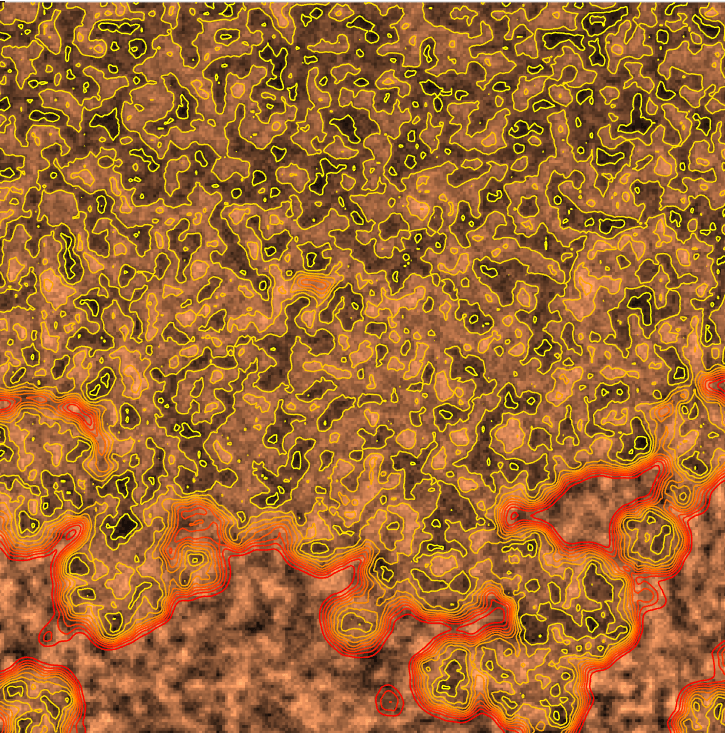}}
	\subfigure[colorbars]{\raisebox{-0.3em}{\includegraphics[height=11.65em]{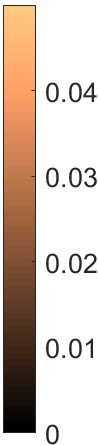}}
		\raisebox{-0.12em}{\includegraphics[height=11.4em]{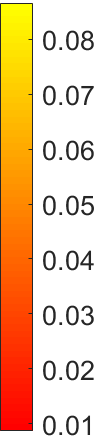}}}
	
	\caption{Experiment \ref{exp:exp.7} (ECM invasion). Shown here the time evolution of the ECM (landscape) and the EC (isolines) over the domain $\Omega =[-2,2]^2$. (a): An initial uniform density of the ECs evolves according to the system \eqref{eq:dcc}--\eqref{eq:ecm}, and mostly proliferates rather than diffuses. (b): The MC particles that are produced by the EMT (not shown here) escape the main body of the tumour, invade the ECM more freely than the EC density, undergo MET and eventually give rise to new EC ``islands''. (d)-(e): These ``islands'' grow mostly due to proliferation and eventually merge with the main body of the tumor. (f): The colorbars for the ECM (left) and the EC density (right) are common to all figures.}\label{fig:exp.7}
	\end{figure}

\end{experiment}

%------------------------------------------------------------------------------------------------------
\section{Discussion}

We have proposed in this work a new modelling approach to study the combined invasion of the ECM by two types of cancer cells, the ECs and the MCs. The proposed framework is a multiscale hybrid model that treats the ECs in a macroscopic and deterministic manner and the MCs in an atomistic and stochastic way. 

We assume that the MCs are much fewer than the ECs and that they emanate from the ECs via a dynamic EMT-like cellular differentiation program. We also assume that the MCs give rise, via the opposite MET-like cellular program, to ECs; this is a key property in the metastasis of the tumour. For simplicity, we assume that both types of cancer cells perform a biased random motion, and that the MCs are much more ``aggressive'' in their migration than the ECs.
 
We encode this information through a hybrid approach: the spatiotemporal evolution of the ECs, the ECM, and the rest of the environmental components are dictated by a macroscopic deterministic model, \eqref{eq:dcc}--\eqref{eq:ecm}. The MCs on the other hand are considered as separate particles that evolve according to a system of SDEs, \eqref{eq:brownian.1}. The transition between the two types of cancer cells is conducted by the \textit{density-to-particles} and \textit{particles-to-density} operators given in \eqref{eq:inter.fwd} and \eqref{eq:inter.bwd}.

This new modelling approach allows us to reproduce in a very natural way, several biological relevant phenomena encountered in the invasion of cancer that are not easily addressed with the usual modelling approaches. Our focus though in this work lies with the description and the handling of the mathematical model and the numerical method; we only present here basic biological situations and postpone the more elaborate investigations for a follow-up work. 

With the atomistic component of our model, we are able to reproduce a sustainable invasion of the ECM by means of a \textit{self-induced haptotaxis gradient} as shown in Experiment \ref{exp:exp.8}. Such behaviour is observed in biological situations and becomes crucial to several biological processes like \textit{wound healing}. The detailed study of such cases falls beyond the scope of the current paper; here we use this experiment as an indication that our model can reproduce biologically relevant situations.

With the full model, we are able to reproduce the spread of the tumour and the invasion of the ECM in the form of invasion ``islands'', Experiment \ref{exp:exp.7}. These are well known to appear in many cases of cancer and are quite challenging to reproduce by either macroscopic or atomistic cancer invasion models. With our approach these invasion ``islands'' appear in a very natural way, and ---most notably--- they appear outside the main body of the tumour. 

What is also very natural in our approach, is that the MC particles escape the main body of the tumour and \textit{remain undetected} while they invade the ECM. It is only after they have established new ``islands'' in the vicinity of the original tumour or in another location within the organism that they can be detected. This is another sought-after property in the field of cancer invasion modelling. 

%From a mathematical point of view, further extensions of this work should include the analysis of the hybrid model; among others the \textit{existence}, \textit{uniqueness}, \textit{boundedness}, \textit{positivity}, and \textit{long time asymptotic} behaviour of the solutions should be addressed. Moreover, rigorous numerical analysis investigations of the \textit{stability}, the  \textit{consistency}, and the \textit{convergence} of the combined numerical method should also be conducted.

For the sake of presentation, we have only considered here some of the fundamental properties of cancer growth that our model can reproduce. Still they suffice to warrant extensions and investigations of more realistic biological situations and experimental settings. To mention but a few: extension to three-dimensional space, more realistic EMT and MET transitions, interactions between cancer cells of the same and different type including collisions, adhesions, short or long range interactions, and the \textit{collective behaviour} of cancer cells.

%---------------------------------------------------------------
\section*{Data Management}
All the computational data output is included in the present manuscript. 

%---------------------------------------------------------------
\section*{Acknowledgements}
NS was partly funded from the German Science Foundation (DFG) under the grant SFB 873:  ``Maintenance and Differentiation of Stem Cells in Development and Disease''. For the work AM is partly supported by funding from the European Union Horizon 2020 research and innovation programme under the Marie Sk\l{}odowska-Curie grant agreement No 642866, the  Commission for  Developing  Countries,  and was partially supported by a grant from the Simons Foundation. AM is a Royal Society Wolfson Research Merit Award Holder funded generously by the Wolfson Foundation. The authors (NS, AM, MC) would also like to thank the Isaac Newton Institute for Mathematical Sciences for its hospitality during the programme [Coupling Geometric PDEs with Physics for Cell Morphology, Motility and Pattern Formation] supported by EPSRC Grant Number EP/K032208/1. 

%---------------------------------------------------------------
%\section*{Supplementary Material}
%Two simulation that correspond to Experiments \ref{exp:exp.8} and \ref{exp:exp.7}, i.e. \texttt{Self\textunderscore generated\textunderscore gradient}  and \texttt{ECM\textunderscore invasion}  respectively.
%------------------------------------------------------------------------------------------------------
%------------------------------------------------------------------------------------------------------
%\begin{multicols}{2}
	\small
	\bibliographystyle{plain}%{alpha,plain, unsrtnat}
	\bibliography{hybrid_model}
%\end{multicols}

%------------------------------------------------------------------------------------------------------
%------------------------------------------------------------------------------------------------------
\appendix 
\small 
%------------------------------------------------------------------------------------------------------

%---------------------------------------------------------------
\section{(Numerical method for the ARD model \eqref{eq:w.model})}\label{sec:FV}
We use a second order \textit{Implicit-Explicit Runge-Kutta} (IMEX-RK) \textit{Finite Volume} (FV) numerical method that was previously developed in \cite{Sfakianakis.2016,Sfakianakis.2016b} where we refer for more details, see also \cite{Madzvamuse.2012}. Here we provide some basic description of the method.

We consider a generic ARD system of the form 
\begin{equation}\label{eq:app.gen.PDE}
\vec w_t = A(\vec w) + R(\vec w) + D(\vec w),
\end{equation}
where $\vec w$ represents the solution vector, and $A$, $R$, and $D$ the \textit{advection}, \textit{reaction}, and \textit{diffusion} operators respectively. 

We denote by $\vec w_h(t)$ the corresponding (semi-)discrete numerical approximation ---indexed here by the maximal spatial grid diameter $h$---  that satisfies the system of ODEs
\begin{equation}\label{eq:app.gen.scheme}
\partial_t \vec w_h = \mathcal A(\vec w_h) + \mathcal R(\vec w_h) + \mathcal D(\vec w_h),
\end{equation}
where the numerical operators $\mathcal A$, $\mathcal R$, and $\mathcal D$ are \textit{discrete approximations} of the operators $A$, $R$, and $D$ in \eqref{eq:app.gen.PDE} respectively. 

Our method of choice for solving \eqref{eq:app.gen.scheme} is an \textit{Implicit-Explicit Runge-Kutta} (IMEX-RK) method based on a \textit{splitting} in explicit and implicit terms in the form
\begin{equation}\label{eq:app.gen.IMEX}
\partial_t \vec w_h = \mathcal I(\vec w_h) + \mathcal E(\vec w_h).
\end{equation}
The actual splitting depends on the particular problem in hand but in a typical case, the advection terms $\mathcal A$ are treated explicitly in time, the diffusion terms $\mathcal D$ implicitly, and the reaction terms $\mathcal R$ partly explicit and partly implicit.

More precisely, we employ a diagonally implicit RK method for the implicit part, and an explicit RK for the explicit part
%\begin{eqnarray}
%\begin{equation*}%\label{F:eq:IMEX3}
%\begin{array}{l}
%	\vec w_h^{n+1}=\vec w_h^n +\tau_n\Big( \sum\limits_{j=1}^{i-1}b_{j}^E ({\rm -A + R_\text{expl}})(t_n + c_j^E \tau, \vec W_i)
%+\sum\limits_{j=1}^{i}b_{j}^I({\rm D + R_\text{impl}})(\vec W_i) \Big),\\
%\vec W_i= \vec w_h^n +\tau_n \Big( \sum\limits_{j=1}^{i-1}a_{ij}^E ({\rm -A + R_\text{expl}})(t_n + c_j^E \tau, \vec W_i)
%+\sum\limits_{j=1}^{i}a_{ij}^I(\rm D + R_\text{impl})(\vec W_i) \Big) . 
%\end{array}
%\end{equation*}		
%\end{eqnarray}
\begin{equation}\label{eq:app.IMEXRK}
	\left\{	\begin{aligned}
%\left\{\begin{array}{ll}
	&\vec W_i^\ast = \vec w_h^n + \tau_n \sum_{j=1}^{i-2}\bar a_{i,j}\vec E_j + \tau_n \bar a_{i,i-1}\vec E_{i-1},\quad &&i=1\dots s\\
	&\vec W_i = \vec W_i^\ast + \tau_n \sum_{j=1}^{i-1} a_{i,j}\vec I_j + \tau_n a_{i,i}\vec I_i,\quad &&i=1\dots s\\
	&\vec w_h^{n+1} = \vec w_h^n + \tau_n \sum _{i=1}^s\bar b_i \vec E_i + \tau_n \sum_{i=1}^sb_i\vec I_i
%\end{array}\right.,
	\end{aligned}\right.\,,
\end{equation}
where $s=4$ are the stages of the IMEX method, $\vec E_i=\mathcal E(\vec W_i)$, $I_i=\mathcal I(\vec W_i)$, $i=1\dots s$, $\{\bar b,\, \bar A\}$, $\{b,\, A\}$ are respectively the coefficients for the explicit and the implicit part of the scheme, given by the Butcher Tableau in Table \ref{F:tbl:IMEX},  \cite{christopher2001additive}. We solve the linear systems in \eqref{eq:app.IMEXRK} using the \textit{iterative biconjugate gradient stabilised Krylov subspace} method \cite{Krylov.1931, vdVorst.1992}.

\begin{table}[t]
	\caption{Butcher tableaux for the explicit (upper) and the implicit (lower) parts of the third order IMEX scheme \eqref{eq:app.IMEXRK}, see also \cite{christopher2001additive}.}
	\label{F:tbl:IMEX}
	\centering
	\scriptsize
	\begin{tabular}{c|cccc}
		$0$&&&&\\[0.5em]
		$\frac{1767732205903}{2027836641118}$&$\frac{1767732205903}{2027836641118}$&&&\\[0.5em]
		$\frac{3}{5}$&$\frac{5535828885825}{10492691773637}$&$\frac{788022342437}{10882634858940}$&&\\[0.5em]
		$1$&$\frac{6485989280629}{16251701735622}$&$-\frac{4246266847089}{9704473918619}$&$\frac{10755448449292}{10357097424841}$&\\[0.5em]
		\hline\\[-0.5em]
		&$\frac{1471266399579}{7840856788654}$ & $-\frac{4482444167858}{7529755066697}$ & $\frac{11266239266428}{11593286722821}$ & $\frac{1767732205903}{4055673282236}$ 
	\end{tabular}
	%	\\[0.5em]
	%	\footnotesize{\textit{(a):}}
	%\caption{Butcher tableau for the explicit part of a third order IMEX scheme, \cite{christopher2001additive}.}
	%\label{F:tbl:IMEX3expl}
	%\end{table}
	\\[0.5em]
	%%\renewcommand{\arraystretch}{1.5}
	%\begin{table}[t]
	\begin{tabular}{c|cccc}
		$0$&0&&&\\[0.5em]
		$\frac{1767732205903}{2027836641118}$&$\frac{1767732205903}{4055673282236}$&$\frac{1767732205903}{4055673282236}$&&\\[0.5em]
		$\frac{3}{5}$&$\frac{2746238789719}{10658868560708}$&$-\frac{640167445237}{6845629431997}$&$\frac{1767732205903}{4055673282236}$&\\[0.5em]
		$1$&$\frac{1471266399579}{7840856788654}$&$-\frac{4482444167858}{7529755066697}$&$\frac{11266239266428}{11593286722821}$&$\frac{1767732205903}{4055673282236}$\\[0.5em]
		\hline\\[-0.5em]
		&$\frac{1471266399579}{7840856788654}$ & $-\frac{4482444167858}{7529755066697}$ & $\frac{11266239266428}{11593286722821}$ & $\frac{1767732205903}{4055673282236}$ 
	\end{tabular}
	%	\\[0.5em]
	%	\footnotesize{\textit{(b):}}
\end{table}

%---------------------------------------------------------------
\section{(An explicit numerical scheme for the SDE \eqref{eq:brownian.1})}\label{sec:brownian}
We consider an Ito process $X=\{X_t, t_0\leq t\leq T\}$ that satisfies the \textit{geometric Brownian motion} SDE
\begin{equation}\label{eq:brownian}
d\vec X_t = \alpha \vec X_t dt +\beta \vec X_t d\vec W_t,
\end{equation}
where $\vec X_t$ denotes the position in space, and where $\alpha\in \R$ and $\beta>0$ are constants.

We discretise \eqref{eq:brownian} with the explicit Euler-Maruyama scheme as 
\begin{equation}\label{eq:sde.scheme1}
\vec X_{n+1} = \vec X_n + \alpha \vec X_n\tau + \beta \vec X_n \Delta \vec W_t.
\end{equation}
By setting $\Delta \vec W_t =\vec Z\sqrt{\tau}$ with $\vec Z\sim \vec N(0,1)$, \eqref{eq:sde.scheme1} reads 
\begin{equation}\label{eq:sde.scheme2}
\vec X_{n+1} = \vec X_n + \alpha \vec X_n\tau +\beta \vec X_n \vec Z\sqrt{\tau}
\end{equation}
which is a simpler version of the It\^o-type scheme that we employ in \eqref{eq:part.loc}.

For further details on the numerical treatment of \eqref{eq:brownian} and other SDEs we refer to \cite{Iacus.2008, Kloeden.1992}.

%-------------------------------------------------------------------------------------
%\section{A computation to include, or not}
%For every step (even half step) $t^n \rightarrow t^n + \tau =t^{n+1}$, the reaction operator
%\begin{align}
%	A:&\ \frac{c^{n+1} -c^n}{\tau} = - \mu c^n + \rho c^n (1-c^n -v)\\
%	B:&\ \frac{\tilde c -c^n}{\tau} =  \rho c^n (1-c^n -v)\\
%	  &\ \hat c^{n+1} = \tilde c^n - \tau \mu c^n
%\end{align}
%from which we can see that
%\begin{align}
%		\tilde c &= c^{n+1/2}+\frac{\tau}{2}\mu c^n\\
%		\hat c^{n+1} & =  c^{n+1/2}
%\end{align}
%Hence $A$ and $B$ coincide as long as the EMT rate is multiplied by $\tau$ in the split form. 
%
%In the other hand, the diffusion operator is identical between $A$ and $B$.
%\begin{align}
%	A:&\ \frac{c^{n+1} - c^{n}}{\tau} = \Delta c^{n,n+1}\\
%	B:&\ \frac{c^{n+1} - c^{n}}{\tau} = \Delta c^{n,n+1}\\
%\end{align}
%
%The splitting in  $A$ would say $D_{\frac{\tau}{2}} R_{\tau} D_{\frac{\tau}{2}}$, and the difference by $B$ is that there the reaction operator is once again split.

\end{document}